\documentclass[aps,prd,amsmath,superscriptaddress,amssymb,showpacs,floatfix,nofootinbib,longbibliography]{revtex4}
\pdfoutput=1
\usepackage{graphicx}
\usepackage{bm}
\usepackage{times}
\usepackage[dvipsnames]{xcolor}
\usepackage{hyperref}
\hypersetup{
    colorlinks = true,
    linkcolor = {MidnightBlue},
    citecolor = {MidnightBlue},
    urlcolor = {MidnightBlue},
}
\usepackage{slashed}
\usepackage{color}
\usepackage{slashed}
\usepackage{lipsum}
\usepackage{subfigure}
\usepackage{multirow}
\usepackage{amsmath}
\usepackage{array} 
\usepackage{varwidth} 

\newcommand{\be}{\begin{equation}}
\newcommand{\ee}{\end{equation}}
\newcommand{\bea}{\begin{eqnarray}}
\newcommand{\eea}{\end{eqnarray}}

\begin{document}

\title{Simulation and Analysis of Two Toy Models}
\author{Yifan Zhang}
\email{mikehh2021@163.com}
\affiliation{International Department, Beijing ETown Academy, Beijing 100176, P.R. China}
\author{Qing Wang}
\email{wangq@tsinghua.edu.cn }
\affiliation{Department of Physics, Tsinghua University, Beijing 100084, P.R. China}

\begin{abstract}
The matching problem and the distribution law of Galton boards with interactions are studied in this paper. The general matching problem appeals at many scenarios, such as the reaction rate of molecules and the hailing rate of ride-hailing drivers. The Galton board is often used in the classroom as a demonstration experiment for the probability distribution of independent events. The two problems are mathematically modeled and numerically simulated. The expected value of matching rate is derived as an analytical solution of the partial differential equation and confirmed by simulation experiments. The interactions were introduced to Galton boards via two parameters in the toy model, which lead to Gaussian distributions of independent events cannot fit the experimental data well. Instead, 'quantum' Fermi-Dirac distributions unexpectedly conforms to simulation experiments. The exclusivity between particles leads to negative Chemical potential in the distribution function, and the temperature parameter increases with the interaction intensity $\alpha$ and flow rate $N_{sm}$. The relations between parameters can be expressed as a conjecture formula within large parameters range.

\end{abstract}

\maketitle

\section{Modeling and simulating to matching problems}
The general matching problem appeals at many scenarios, such as the reaction rate of molecules and the hailing rate of ride-hailing drivers. 
In this article, we use ride-hailing as an example to discuss matching problems. Simplified model consists of $s$ ride-hailing drivers and $u$ passengers randomly distributed within a unit square. If the ride hailing platform sets the matching distance to $r$, the single probability of any pair of drivers and passengers achieving a match is $p\simeq\pi r^2(1-r)$, here the first-order edge effect is considered. Matching is exclusive, each driver can only match one passenger, and vice versa. The number of matches will be a function depends on the number of drivers $s$, passengers $u$, and the single probability $p$.

\begin{figure}[h]
    \centering
    \includegraphics[width=0.40\textwidth]{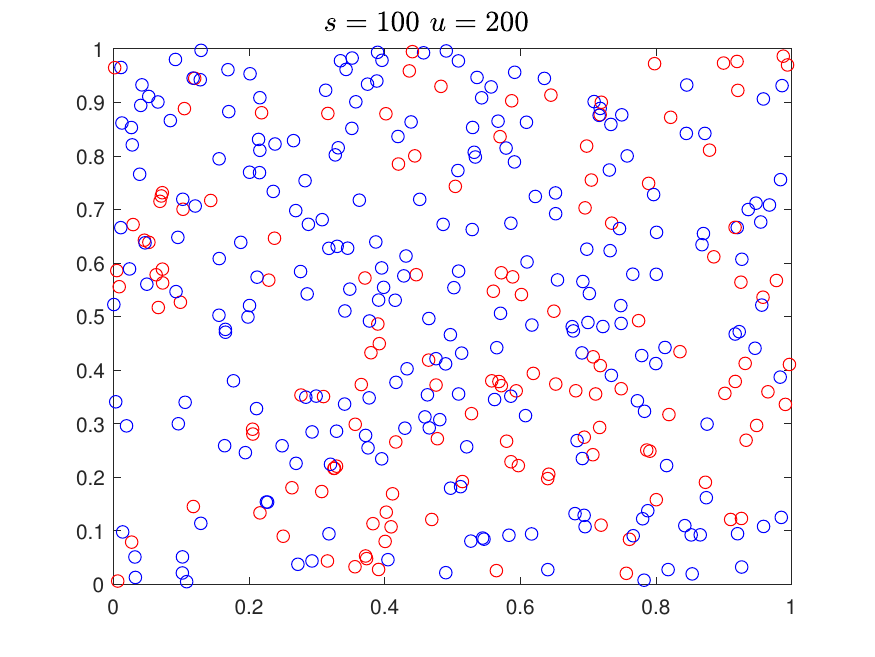}
     \includegraphics[width=0.40\textwidth]{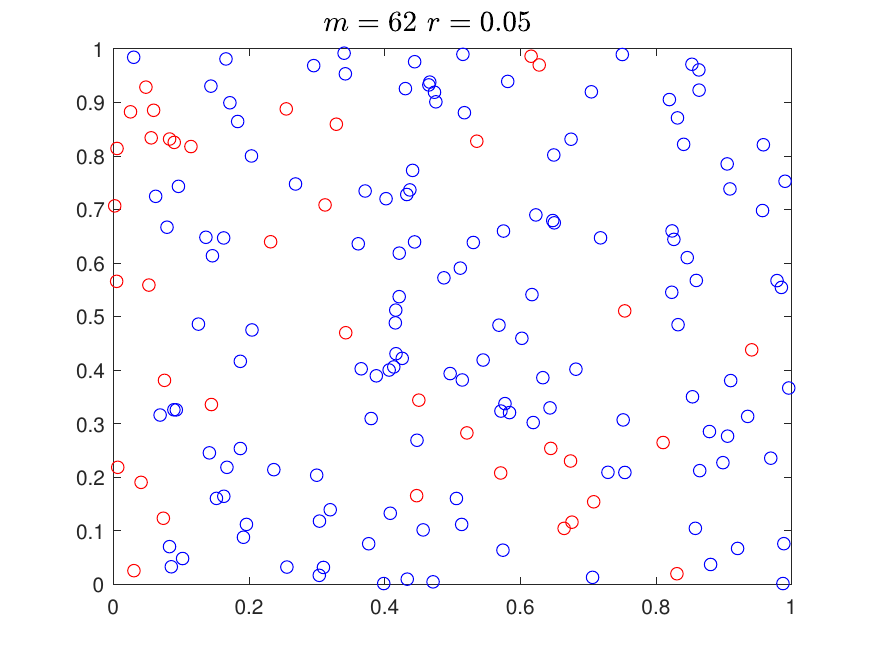}
    \caption{The image of the visual simulation program, where the red and blue circles represent 100 drivers and 200 passengers randomly distributed in the unit square. In this experiment, 62 pairs of drivers and passengers were matched when the matching distance r=0.05.}
    \label{fig: randomus}
\end{figure}

\begin{figure}[h]
    \centering
    \includegraphics[width=0.40\textwidth]{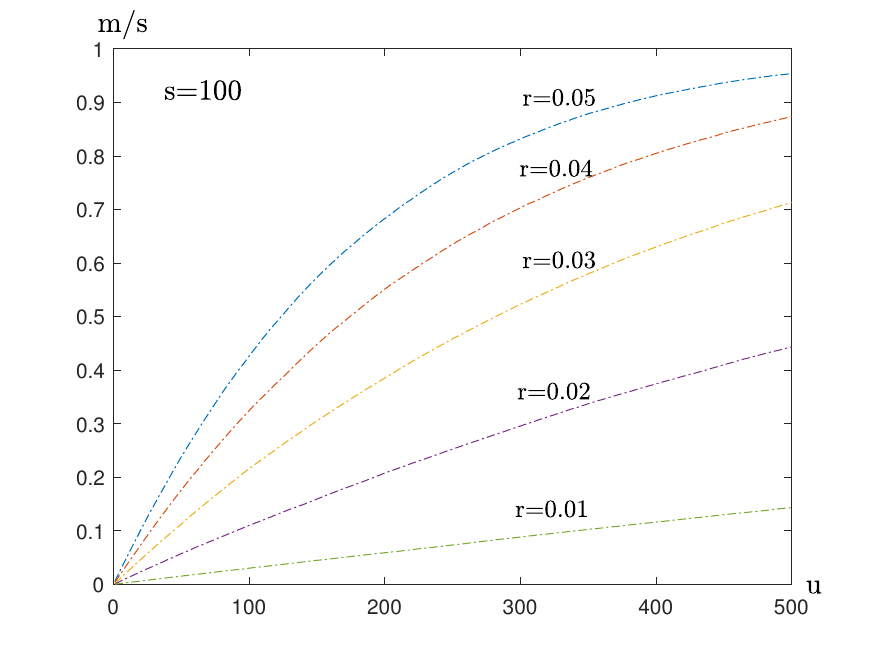}
     \includegraphics[width=0.40\textwidth]{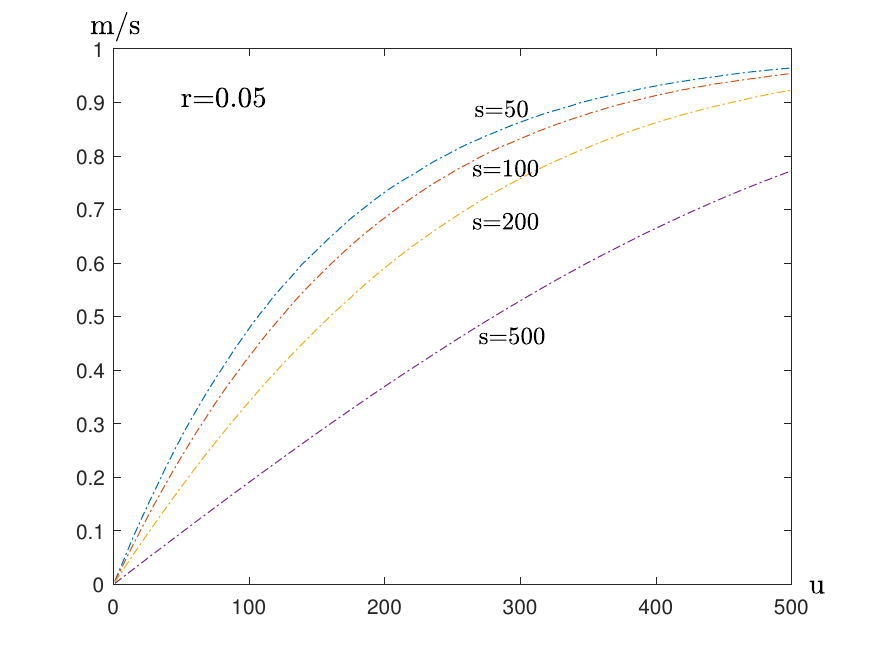}
    \caption{The expected  matching rate for drivers $\bar{m}/s$ obtained by thousands simulation experiments with variable $u$ and different $s$ and $r$.   }
    \label{fig: matchrate}
\end{figure}

The numerical simulation program can generate $s$ two-dimensional random arrays representing drivers' positions and $u$ two-dimensional random arrays representing the passengers' positions. The distance between two sets of random positions is calculated in order of arrangement. Once the distance is less than r, the matched driver and passenger will disappear. After multiple experiments, the expected number of matches can be obtained as shown in Fig.\ref{fig: matchrate}.

\section{Analytic derivation of matching expected function}

The expected value of matching number $\bar{m}$ will be a function depends on the number of drivers $s$, passengers $u$, and the single probability $p$. When both $s$ and $u$ are large($>10$), the function can be derived as an analytical solution of the partial differential equation. If $s$ is fixed and $ps$ is small enough($<10$), the differential equation with the independent variable $u$ and the function value $\bar{m}$ can be established. Increasing $du$ passengers will result in an increase of the expected value $\bar{m}$ by
\begin{equation}
d\bar{m}=(s-\bar{m})p du=\bar{s_L}pdu. \nonumber
\end{equation}
Here, $\bar{s_L}$ is the expected remaining drivers number which satisfies the differential equation 
\begin{equation}
d\bar{s_L}=-\bar{s_L}pdu, \nonumber
\end{equation}
and has solution for fixed s
\begin{equation}
\bar{s_L}=s e^{-pu}. \nonumber
\end{equation}
This means that the remaining number of drivers $\bar{s_L}$ exponentially decreases with the increase of passenger number $u$, with a decay rate of $p$. The expected matching number 
\begin{equation}\label{eq: matching1}
\bar{m}=s-\bar{s_L}= s(1- e^{-pu}). 
\end{equation}
This solution function only fits the simulation experiment when $sp<10$.

Considering the symmetry of $s$ and $u$, the remaining number of passengers $bar{u_L}$ exponentially decreases with $s$
\begin{equation}
\bar{u_L}=u e^{-ps}. \nonumber
\end{equation}
when $u$ is fixed and $pu<10$. The expected matching number 
\begin{equation}\label{eq: matching2}
\bar{m}=u-\bar{u_L}= u(1- e^{-ps}).
\end{equation}

When $s$ and $u$ are both large variables, the expected matching number in \eqref{eq: matching1} and \eqref{eq: matching2} is equal and the ratio of $\bar{s_L}$ and $\bar{u_L}$
\begin{equation}
\frac{\bar{s_L}}{\bar{u_L}}=\frac{s e^{-pu}}{u e^{-ps}}=\frac{s-\bar{m}}{u-\bar{m}}. \nonumber
\end{equation}

The expected matching number has the analytical solution function:
\begin{equation}\label{eq: matching3}
\bar{m}(s,u,p)=\frac{su(e^{pu}-e^{ps})}{u e^{pu}-s e^{ps}}, \,\, \,\,(u\neq s).
\end{equation}
By using the Lopida's law, it can be obtained that
\begin{equation}\label{eq: matching4}
\bar{m}(s,u,p)=\frac{ps^2}{1+ps},\,\, \,\, (u=s).
\end{equation}

\begin{figure}[h]
    \centering
    \includegraphics[width=0.40\textwidth]{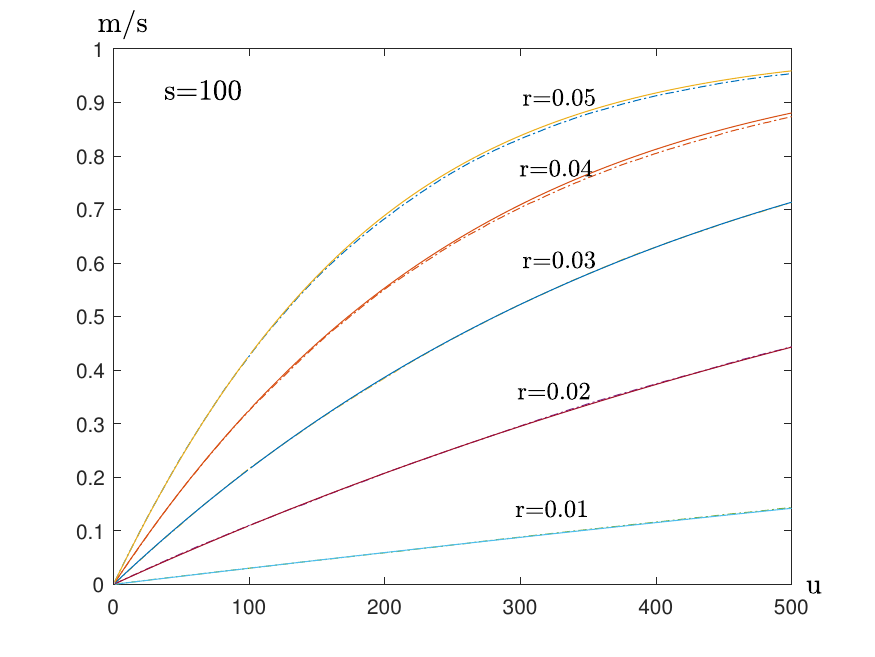}
     \includegraphics[width=0.40\textwidth]{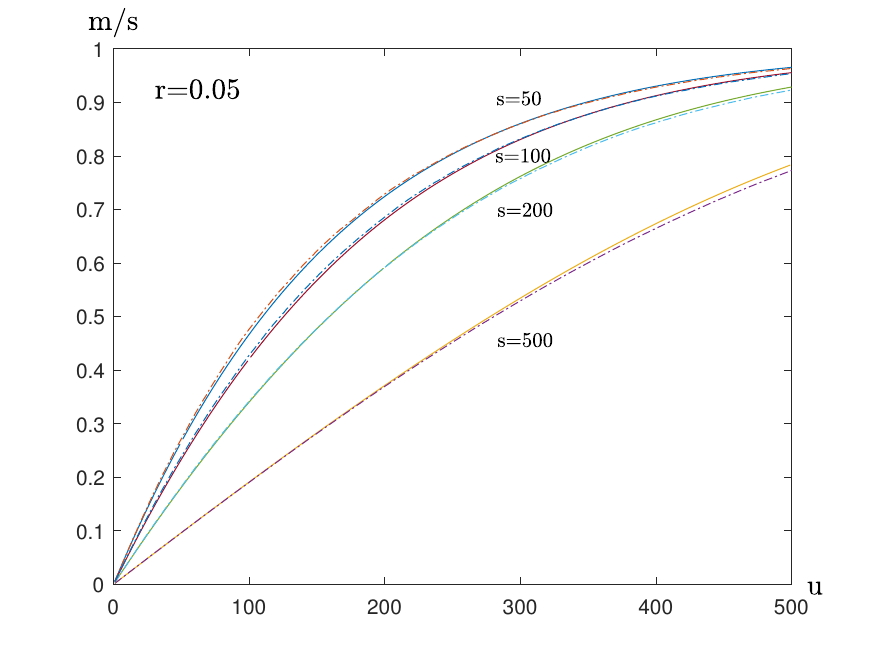}
    \caption{The curves of expected  matching rate for drivers $\bar{m}/s$ in Eq.\ref{eq: matching3}  and \ref{eq: matching4} with variable $u$ and different $s$ and $r$ are consist with the simulation curves. }
    \label{fig: matchanaly}
\end{figure}

The analytical formula is highly consistent with the simulation experiments as shown in Fig.\ref{fig: matchanaly}. The tiny deviations of the curves in the figures come from the second-order effect of the edge (ignored in $p$ value) when r is not small.

\begin{figure}[h]
    \centering
    \includegraphics[width=0.40\textwidth]{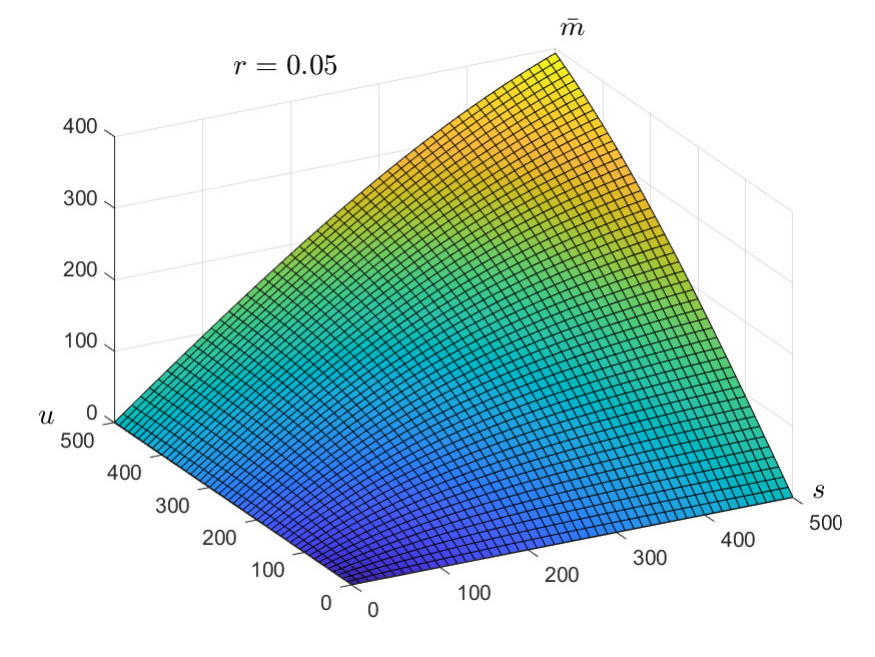}
     \includegraphics[width=0.40\textwidth]{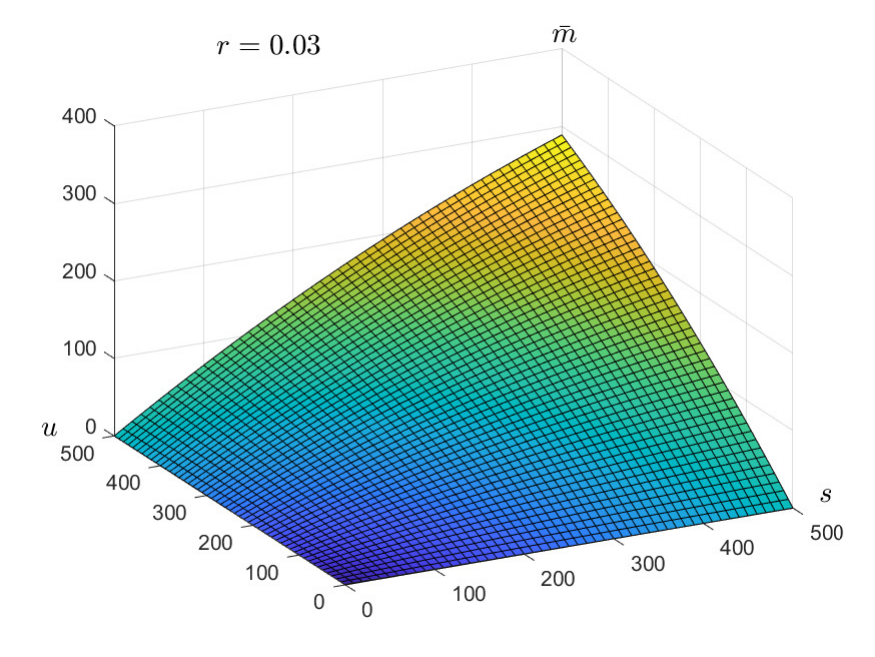}
    \caption{The surfaces of expected  matching number  $\bar{m}$ in Eq.\ref{eq: matching3}  and \ref{eq: matching4} for variable $u$ and  $s$ and different $r$. }
    \label{fig: Fig_mrfurf}
\end{figure}

The matching number $\bar{m}$  can be easily obtained by the effective analytical formula as the surfaces in Fig.\ref{fig: Fig_mrfurf}.  It is convenient to adjust the matching distance $r$ based on the $s$ and $u$ values to control the appropriate matching rate.

\section{Introduction to Galton boards}

The Galton board \cite{galton} is usually used as a toy experiment to demonstrate probability distributions of independent events. The simple model also can study metal electron gas systems after considering dynamic details \cite{Lorentz, Moran,KR}. Besides the studies of physics, Galton boards are also used in medical and traffic research \cite{CS, LL}. We use the following simplified model to discuss the working principle of Galton board. As shown in Fig.~\ref{fig: fig1}, the Galton board device consists of evenly spaced vertical slots below and regularly arranged horizontal bars in the middle. The upper container contains a large amount of particles. Open the small outlet at the bottom of the container, and a large number of particles will flow down from top to bottom, colliding with the horizontal bars of each layer  and falling into the slots. The Galton board simulates the track of particle motion, counts the number of particles falling into the slot and their proportion to the total number of particles.

\begin{figure}[h]
    \centering
    \includegraphics[width=0.40\textwidth]{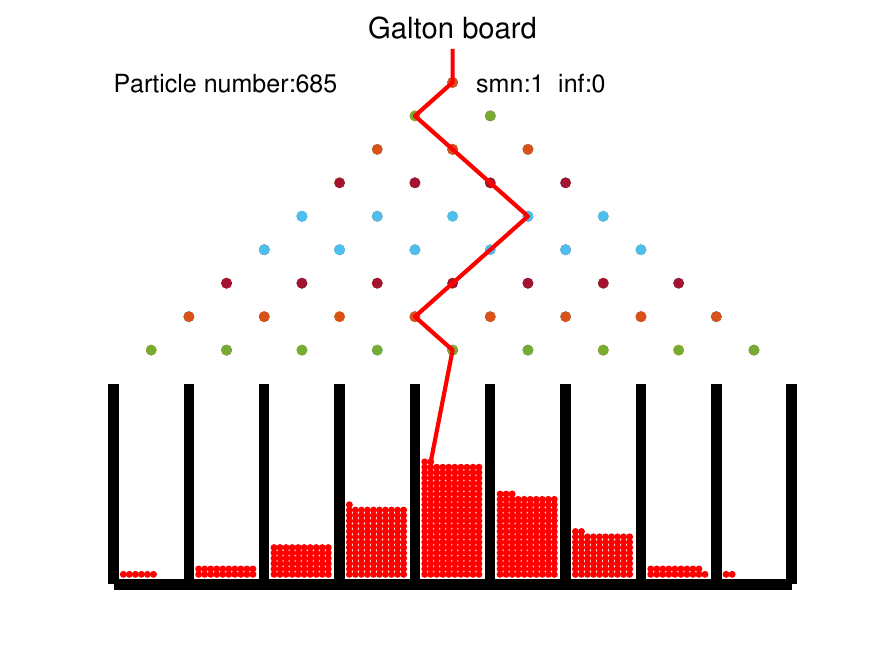}
    \caption{The imagine of particles randomly falling in a Galton board generated by MATLAB. The Galton board consists of 9 slots below and 9-layer horizontal bars(the dots in the figure represent the spaces between bars ). Particles fall one by one, The 685th particle enters 5th slot through a random path as shown in the figure. }
    \label{fig: fig1}
\end{figure}

The Galton board has been easily approached using numerical simulations experiments\cite{kozlov}, with the development of numerical calculation software. Numerical calculations can even calculate the dynamic track of particle falling \cite{Garwin, cross} .  In our simulation experiments, particles move left or right by one grid according to the set probability for each layer they fall, which is equivalent to one binomial selecting distribution event. If particles fall alone or ignore interactions between particles, the probability of left and right shifts is the same. In this case, after passing $m$-layer horizontal bars,  one particle falls into one of $m$ slots and the distribution of position satisfies the binomial distribution:
\begin{equation}\label{eq: binomial}
P^i_m = \frac {C^i_2}{2^m} = \frac {m!}{(i-m)!i!2^m} ,
\end{equation}
Here, $P^i_m$ is the probability of particles falling into the $i$-th slot in  total $m$ slots.

When the number of layers/slots in the Galton board is large enough($m \geq 9$), the distribution is approximately a Gaussian distribution function.

 \begin{equation}\label{eq: Gaussian}
f(x)= \frac {1}{\sqrt{2\pi}\sigma} \exp (-  \frac {x^2}{2\sigma^2}) ,
\end{equation}

Here, $x=(i-m)/2$ is the slot position where particles entered. The standard deviation $\sigma=1.76$ when $m=13$ and  $\sigma=1.45$ when $m=9$ which according to the following theoretical formula
 \begin{equation}\label{eq: deviation}
\sigma= \frac {\sqrt{m-1}}{2} + \frac {1}{\sqrt{2}\pi^2 \ln m} ,
\end{equation}

 We simulated this process by MATLAB and obtained results that are consistent with theory.

\begin{figure}[h]
    \centering
    \includegraphics[width=0.40\textwidth]{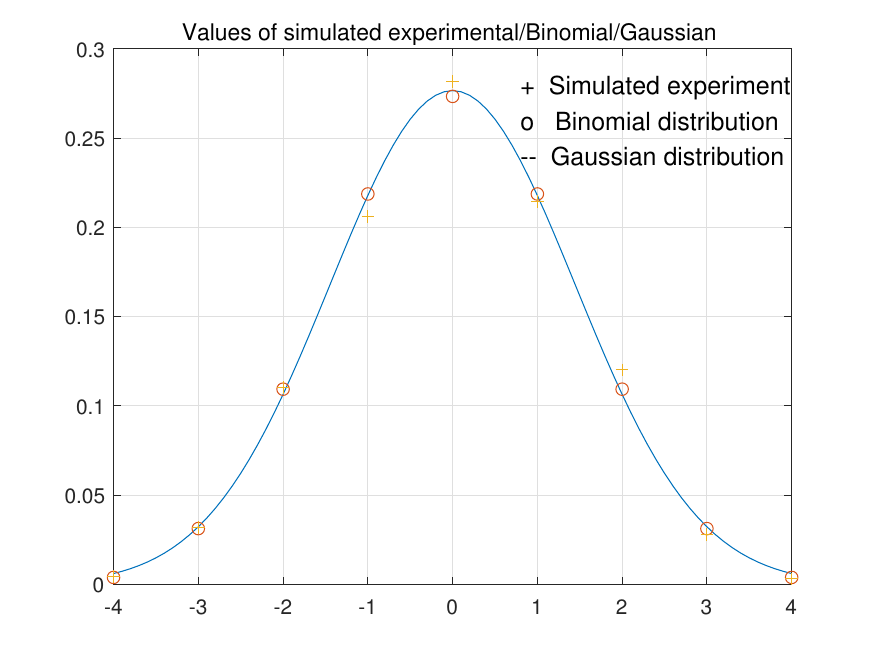}
    \caption{The simulation experiment values and fitted Gaussian distribution function(standard deviation $\sigma=1.45$) compare to the binomial distribution $P^i_m$ as total slots number $m=9$. }
    \label{fig: fig1}
\end{figure}

\section{Introduction and parameterization of interparticle squeezing effects}
\label{sec: methodology}

Introduction and parameterization of interparticle squeezing effects.

The probability of particles moving left or right is influenced by the number of particles on both sides, with fewer particles and more vacancies corresponding to a higher probability. To simplify the problem, we assume that the probability of selecting left or right moving is influenced by the relative difference of particle numbers at two sides. If the number of particles in same grid is $N_M$, on the left is $N_L$, and $N_R$ on the right, then the probability of left shifting will increased by $\Delta p_L = \frac{\alpha}{2} (N_R-N_L)/(N_L+N_M+N_R) $, where the factor $\alpha$ represents the strength of the interaction, which is determined by the ratio of space between horizontal bars to particle diameter. The total probability of left and right shifting is normalized as $ p_L+p_R = 1$. So, the probabilitis

\begin{equation}\label{eq: probability}
p_L=\frac{1}{2} +\frac{\alpha}{2} \frac{N_R-N_L}{N_L+N_M+N_R}
\end{equation}
\begin{equation}
p_R=\frac{1}{2} -\frac{\alpha}{2} \frac{N_R-N_L}{N_L+N_M+N_R}
\end{equation}

\begin{figure}[h]
    \centering
    \includegraphics[width=0.40\textwidth]{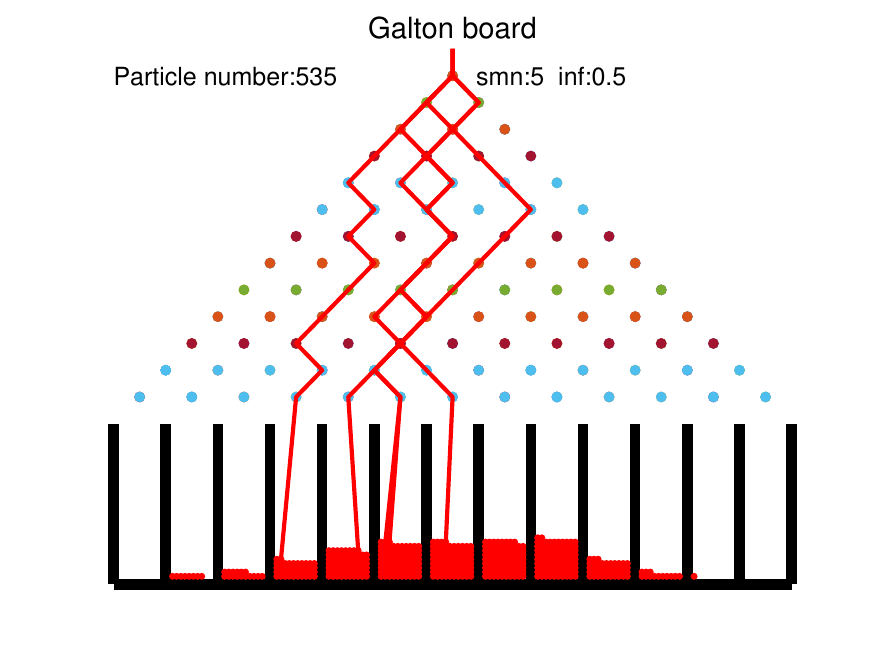}
    \caption{The imagine of particles randomly falling in a Galton board generated by MATLAB. The Galton board consists of 13 slots below and 13-layer horizontal bars. $N_{sm}=$ 5 particles fall simultaneously, interaction factor $\alpha=0.5$. The five particle paths are shown in the figure. }
    \label{fig: fig3}
\end{figure}

We simulated the process of $N_{sm}$ particles falling simultaneously by MATLAB as shown in Fig.~\ref{fig: fig3}.

\begin{figure}[h]
    \centering
    \includegraphics[width=0.4\textwidth]{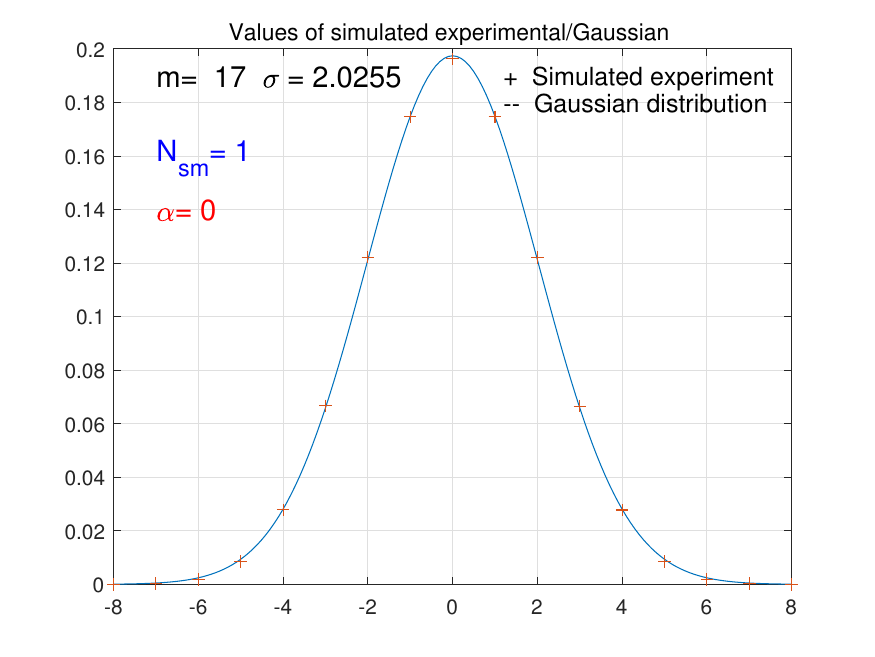}
    \includegraphics[width=0.4\textwidth]{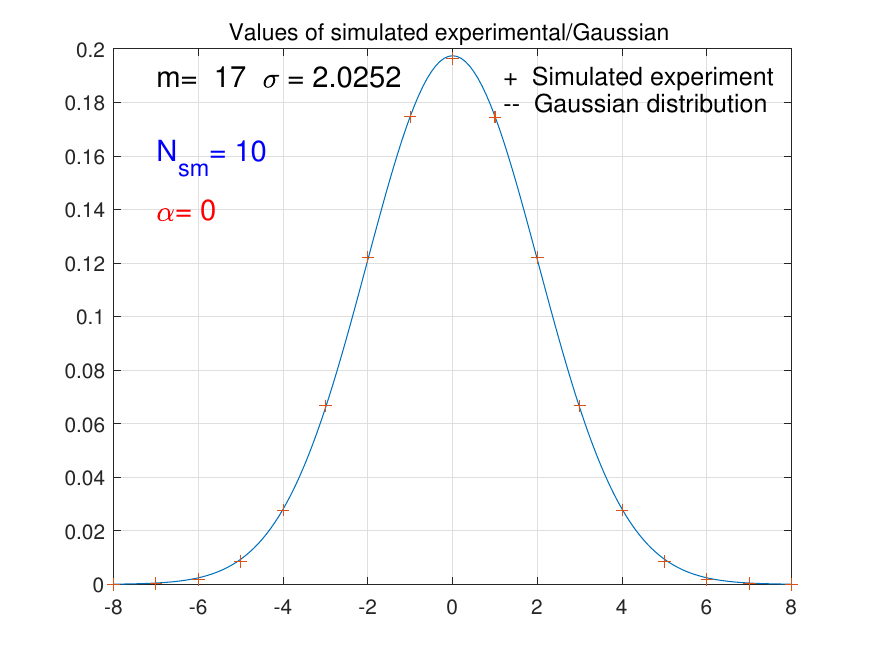}
    \caption{The fitted Gaussian distribution function(standard deviation $\sigma=2.03$) compare to the simulation experiment values as total slots number $m=17$. The consistency of data is better than that in Figure 2 because the layer number m is larger. when $\alpha$ is zero,  the distribution functions of $N_{sm}=1$ (Left) and $ N_{sm}=10$(Right) are the same Gaussian function. }
    \label{fig: fig4}
\end{figure}

If $\alpha$ is zero, this means the particles are tiny and the interaction between particles is negligible. In this case, no matter how many particles fall simultaneously, the probability $p_L=p_R=1/2$ and the distribution function is the same as that of free particle passing through the Galton board. The simulation experiment results verify this conclusion, as shown in Fig.~\ref{fig: fig4}, the simultaneous falling particle number $N_{sm}=1$ and $N_{sm}= 5$, the distribution functions are the same Gaussian function.

\begin{figure}[h]
    \centering
    \includegraphics[width=0.4\textwidth]{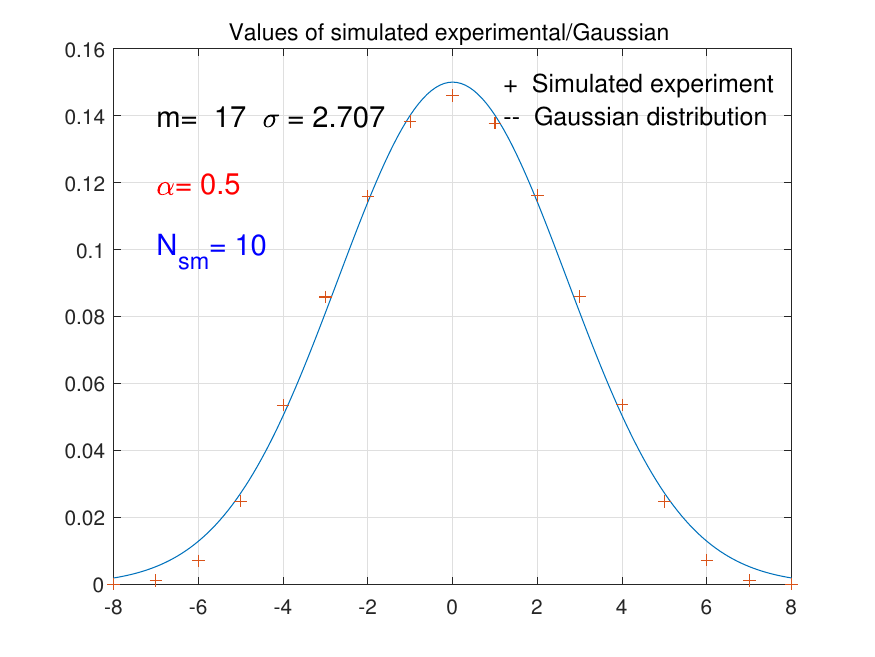}
      \caption{The simulation experiment values and best-fitting Gaussian distribution function(standard deviation $\sigma=2.7$), $ N_{sm}=10$ and $\alpha= 0.5$ . The simulation distribution obviously deviates from the Gaussian function. }
    \label{fig: fig5}
\end{figure}

When both $N_{sm}$ and $\alpha$  increase, the simulation distribution obviously deviates from the Gaussian function and can not fitted by a broaden Gaussian function as shown in Fig.~\ref{fig: fig5}.

\begin{figure}[h]
    \centering
    \includegraphics[width=0.4\textwidth]{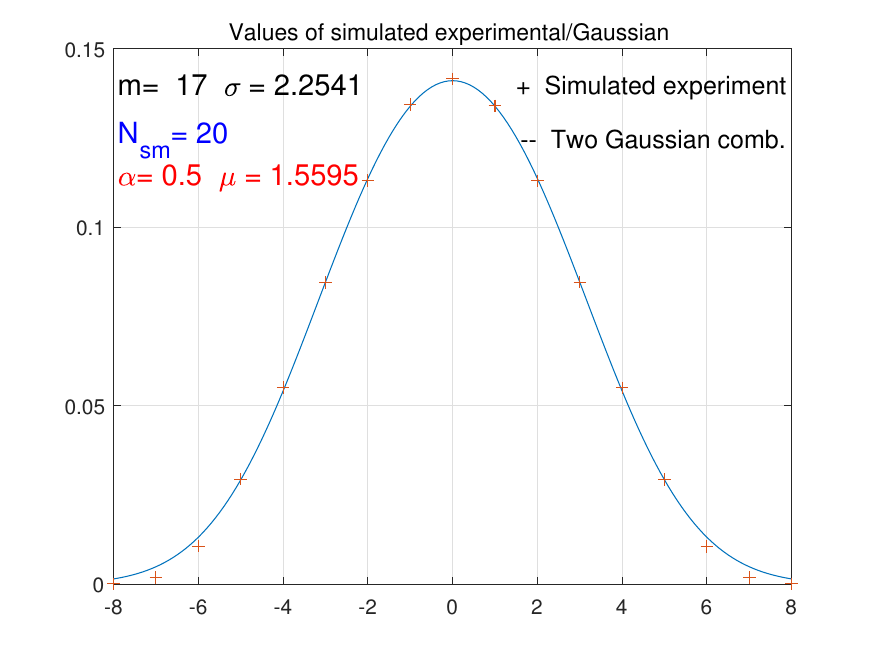}
    \includegraphics[width=0.4\textwidth]{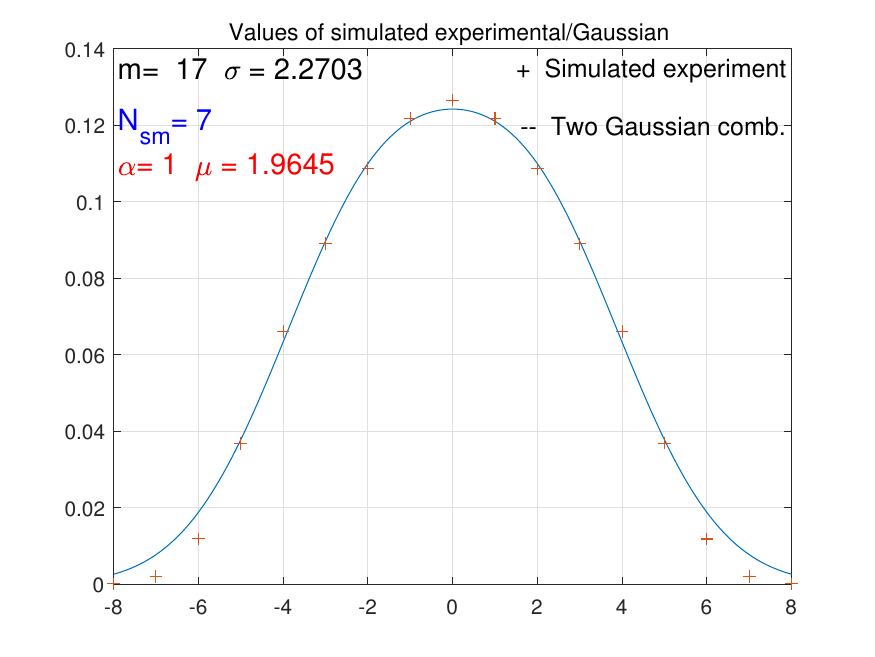}
      \caption{The simulation experiment values and best-fitting shifted Gaussian functions(standard deviation $\sigma=2.25$, mean value $\mu=1.56 $), $ N_{sm}=20$ and $\alpha= 0.5$ (Left).  ($\sigma=2.27$, mean value $\mu=1.96 $), for $ N_{sm}=7$ and $\alpha= 1$(Right) }
    \label{fig: fig6}
\end{figure}

 Firstly,  it is assumed that the result of the simulation experiment is two extended Gaussian distributions with left/right shift in the mean value. We use two symmetric Gaussian functions with the center value shift $\mu$ to fit the simulation experiment results
 \begin{equation}\label{eq: Gaussian2}
f(x)= \frac {1}{2\sqrt{2\pi}\sigma} [\exp (-  \frac {(x+\mu)^2}{2\sigma^2}+\exp (-  \frac {(x-\mu)^2}{2\sigma^2})] ,
\end{equation}
Then, two shift Gaussian functions has a much better fit with than simulation results as shown in Fig.~\ref{fig: fig6} when $N_{sm}$ is large and  $\alpha$ not close to 1. But two  Gaussian fitting is not good when small $N_{sm}$ and $\alpha>0.7$.

\begin{figure}[h]
    \centering
    \includegraphics[width=0.4\textwidth]{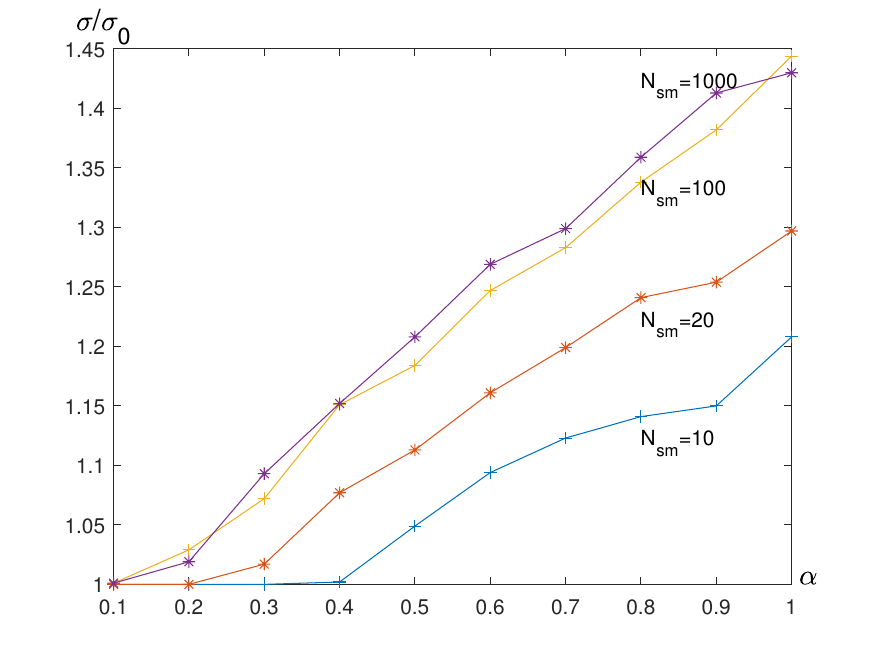}
    \includegraphics[width=0.4\textwidth]{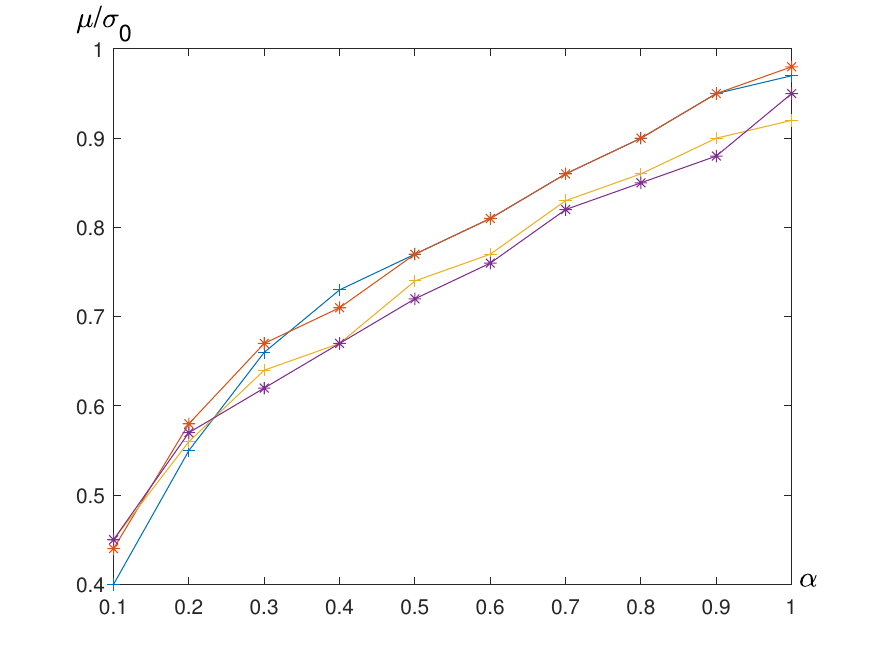}
      \caption{The best-fitting parameters $\mu$ and $\sigma$ vs variant $\alpha$  for different $ N_{sm}$  }
    \label{fig: fig7}
\end{figure}

Meanwhile, the best-fitting parameters $\mu$ and $\sigma$ are presented as complex function curves with the different  $ N_{sm}$ and $\alpha$ as shown in Fig.~\ref{fig: fig7}.

The $\mu$ and $\sigma$ values of the Gaussian function increase with $N_{sm}$ and $\alpha$ parameters. As shown in the Fig.~\ref{fig: fig4}. The results indicates that the number of particles falling simultaneously $N_{sm}$ and the interaction factor $\alpha$ cause the particles to disperse on both sides due to collision repulsion during the falling process.

In summary, the extended Gaussian distribution cannot describe the experimental results.

\subsection{Simulation Models}
\label{sec: simulation models}
Considering the exclusivity caused by interactions and wholeness of particles, we attempt to fit the experiment using the Fermi Dirac distribution function:
\begin{equation}\label{eq: FD_equation}
f(x)\propto \frac {1}{e^ {\frac {E-E_F}{1+b}}+1};\  E= \frac {x^2}{2\sigma^2}.
\end{equation}
The temperature parameter $1+b$ and Fermi energy(Chemical potential) $E_F$ in this function are obtained through fitting experiments.
The exclusivity between particles leads to negative Chemical potential when $N_{sm}$ is larger, and the temperature parameter increases with the interaction intensity $\alpha$ and flow rate $N_{sm}$. Fermi Dirac distributions functions can fit experiments well within the full parameter range for $N_{sm}=2\to \infty$ and $\alpha=0 \to 1$. The best fitting Fermi Dirac distribution curves  compared with experiments are shown in Fig.~\ref{fig: fig8} and Fig.~\ref{fig: fig9}.

\begin{figure}[h]
    \centering
    \includegraphics[width=0.4\textwidth]{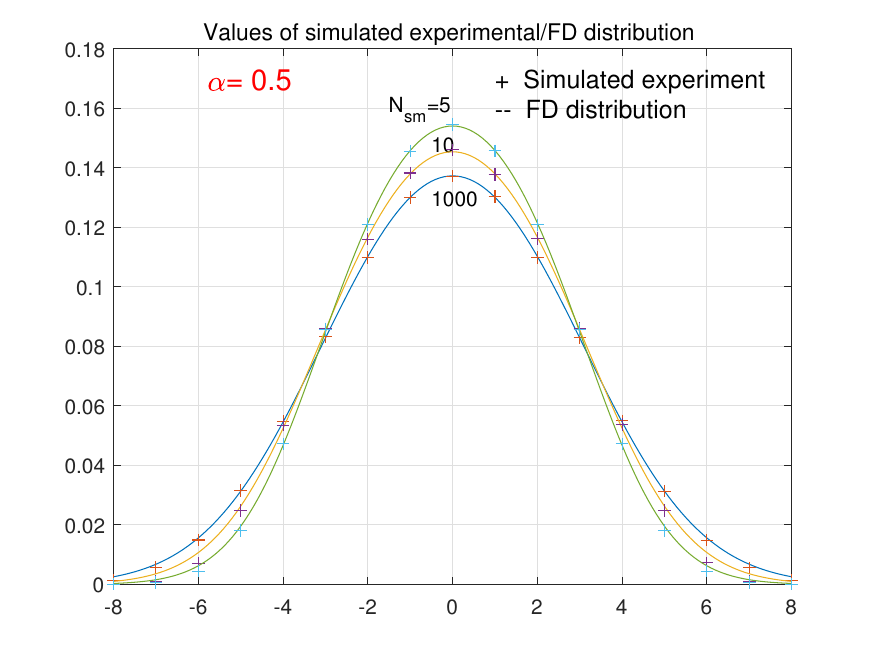}
      \caption{The simulation experiment  with fixed $\alpha=0.5$ and variant $ N_{sm}=5,10,1000 $, compared with the best fitting Fermi Dirac distributions  }
    \label{fig: fig8}
\end{figure}

\begin{figure}[h]
    \centering
    \includegraphics[width=0.4\textwidth]{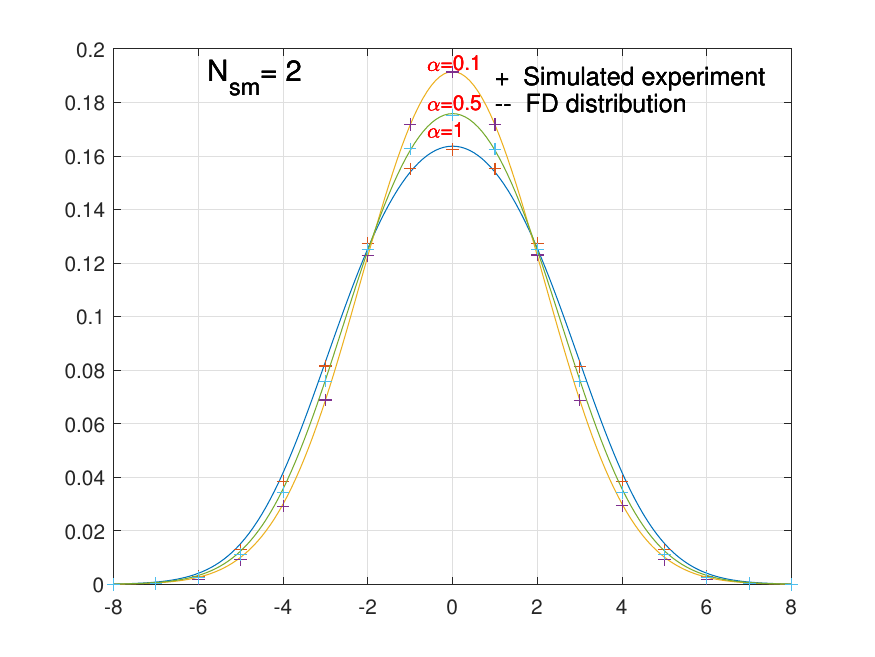}
     \includegraphics[width=0.4\textwidth]{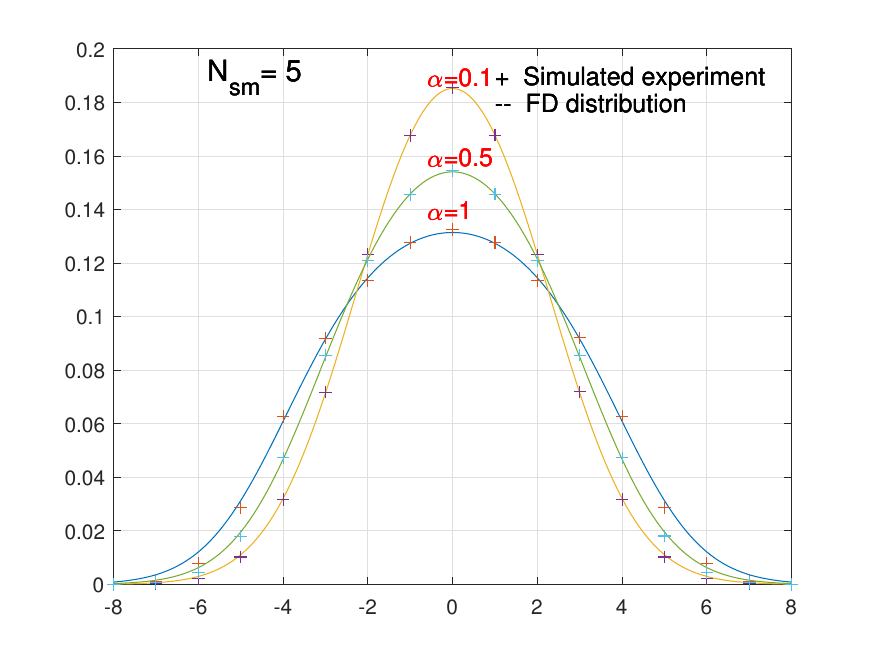}
      \includegraphics[width=0.4\textwidth]{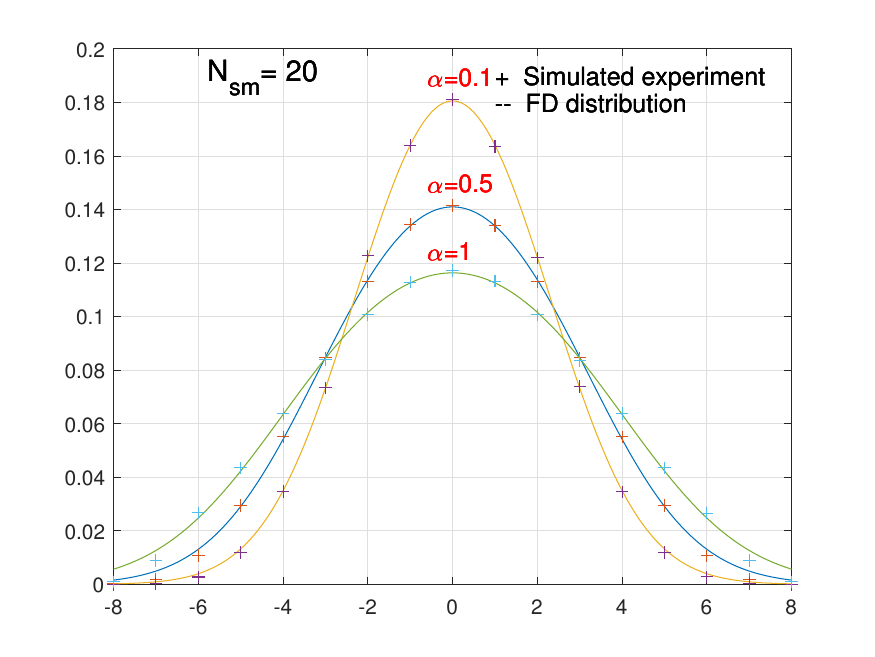}
       \includegraphics[width=0.4\textwidth]{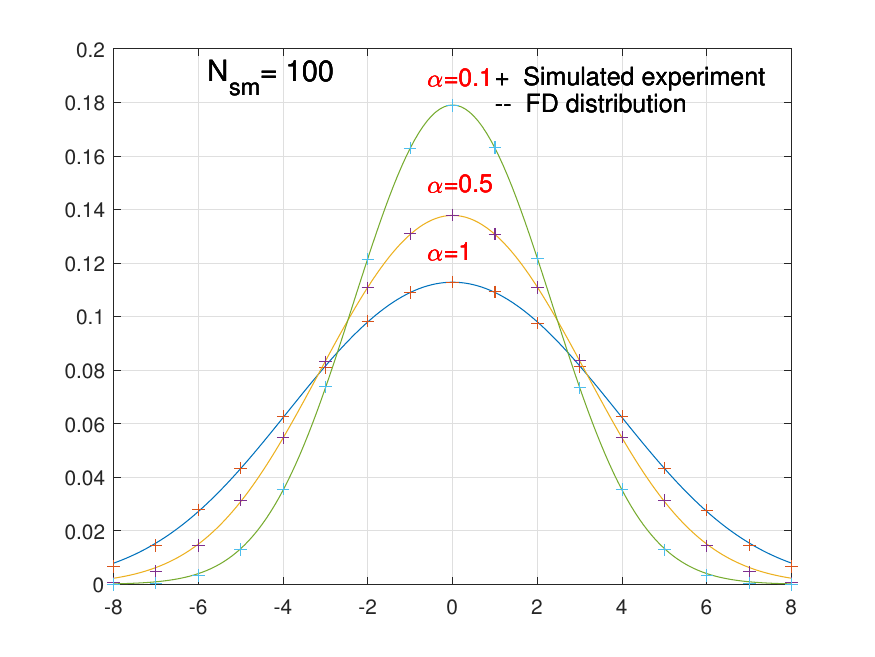}
      \caption{The simulation experiment  with variant $\alpha$ and  $ N_{sm} $, compared with the best fitting Fermi Dirac distributions  }
    \label{fig: fig8}
\end{figure}

\begin{figure}[h]
    \centering
    \includegraphics[width=0.4\textwidth]{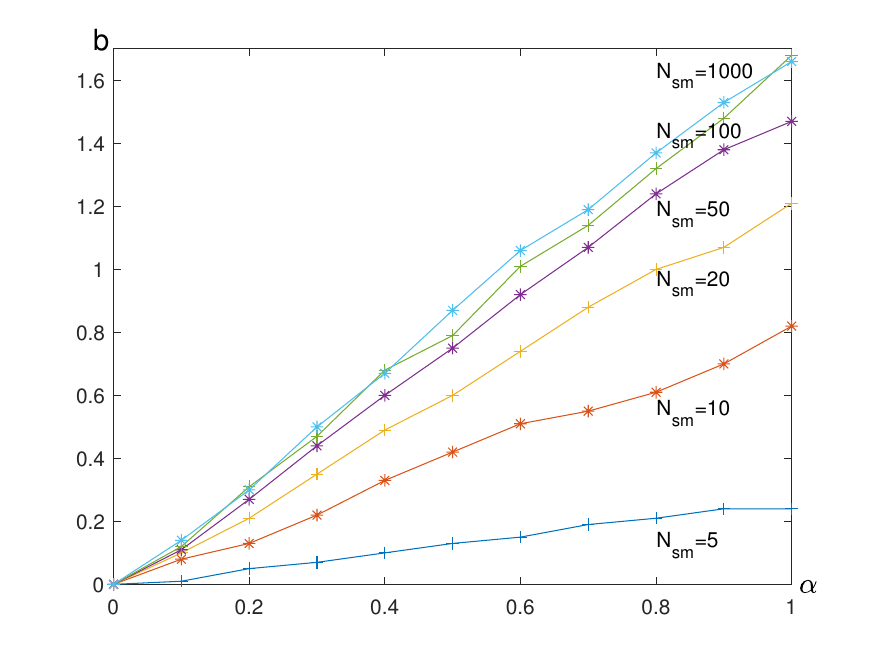}
     \includegraphics[width=0.4\textwidth]{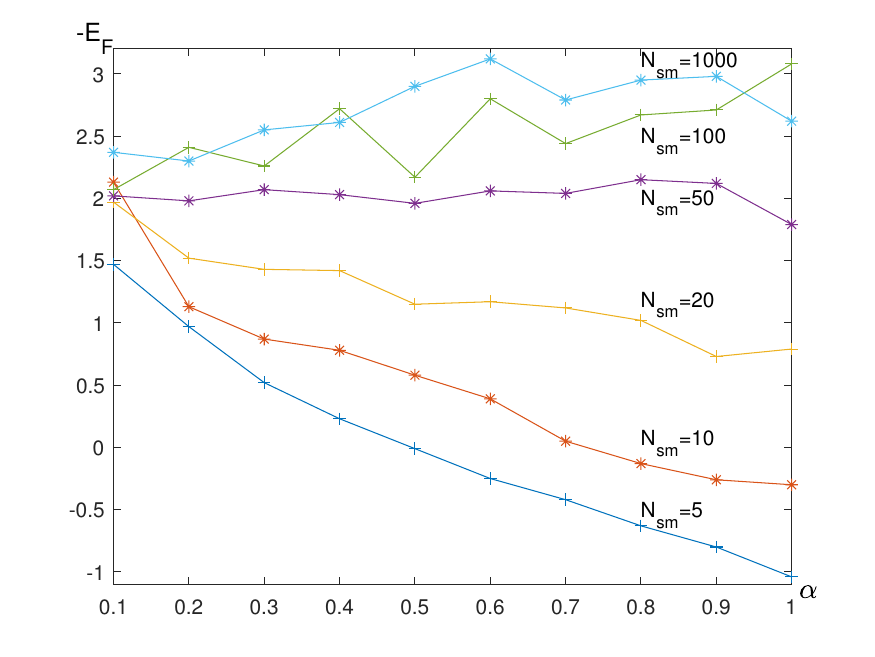}

      \caption{The  the best fitting parameters $b$ and $-E_F$  with variant $\alpha$ and  $ N_{sm} $, Fermi energy $E_F$ have significant fluctuations as $m=17$. }
    \label{fig: fig10}
\end{figure}

Although the Fermi Dirac distribution function can describe experiments well, the fitting parameters especially Fermi energy $E_F$ have significant fluctuations as shown in Fig.~\ref{fig: fig10}. The reason for the fluctuations is that most particles are distributed in the 3$Sigma$ range about 7 sluts, which leads to  the available data for fitting is not enough.

In order to find the relationship between the Fermi Dirac distribution parameters $b$, $-E_F$ and experimental parameters $\alpha$ and  $ N_{sm} $, the more slut-number $m$ is needed. When the slut-number $m=25$,  the fluctuation of $b$ parameter is reduced  significantly.
The parameter $b$ exhibits a simple linear dependence on the interaction intensity $\alpha$, when flow rate $N_{sm}>10$.

\begin{figure}[h]
    \centering
    \includegraphics[width=0.4\textwidth]{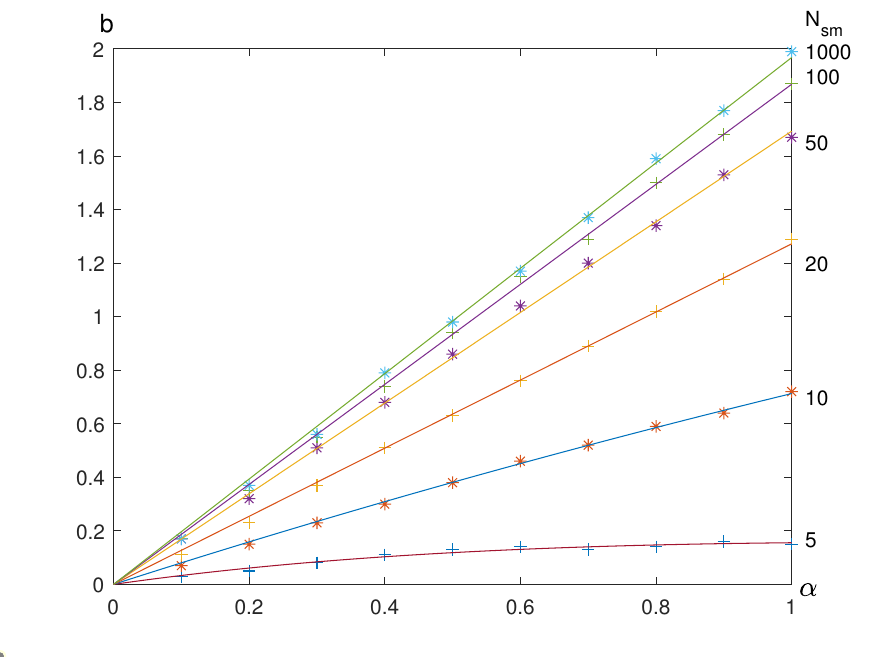}
     \includegraphics[width=0.4\textwidth]{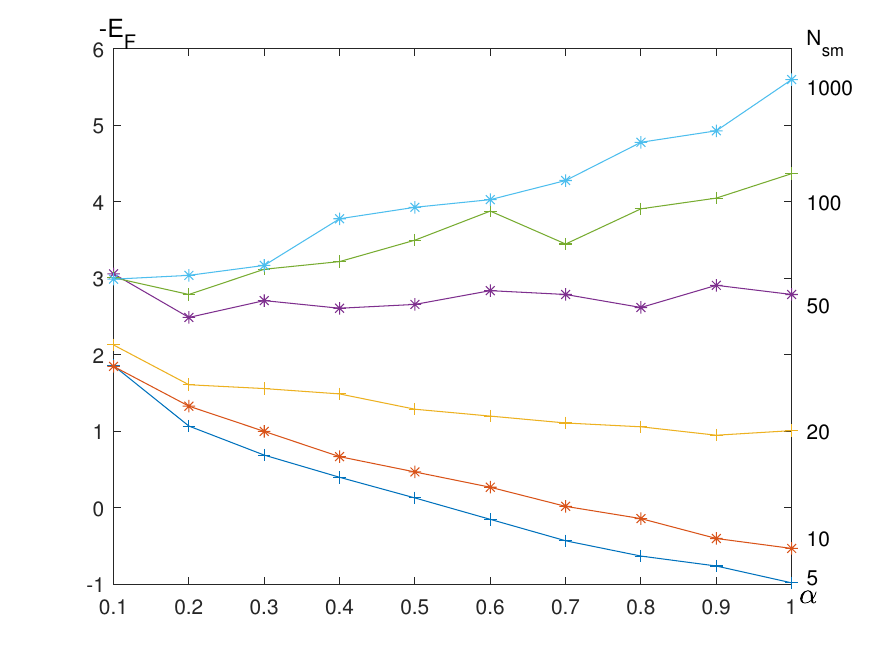}

      \caption{The  the best fitting parameters $b$ and $-E_F$  with variant $\alpha$ and  $ N_{sm} $ as  $m=25$, the parameter $b$ can be simply expressed as a formula $b=\kappa_b \alpha e^{-\eta_b \alpha}$ . }
    \label{fig: fig11}
\end{figure}

The 'temperature' parameter $b$ can be simply expressed as:
\begin{equation}
 b=\kappa_b \alpha e^{-\eta_b \alpha};
\end{equation}

Here, the slop $\kappa_b$ slowly increases with the flow rate $N_{sm}$ as the conjecture formula
\begin{equation}
 \kappa_b= 1.99(1-\gamma^2);\ \ \gamma=\frac{m-1}{N_{sm}+m-2}.
\end{equation}

The exponential factor $\eta_b$ rapidly decreases to zero with increasing $N_{sm}$
\begin{equation}
 \eta_b= 120\beta^{4};\ \ \beta=\frac{1}{N_{sm}-1}
\end{equation}

When $N_{sm}$ is large, Fermi energy $E_F$ is not sensitive to the quality of fitting.  Otherwise, the fitted Fermi energy $E_F$ is not accurate and has large fluctuations as shown in Fig.~\ref{fig: fig11}. It's  need to find a new parameter that is sensitive to fitting. We find that the ratio of the FD function's second derivative to the value of function  at the origin is a good parameter. The new sensitive parameter $\xi$ (the ratio minus 1)is defined as

\begin{equation}
 \xi= b+(1+b) e^{\frac{E_F}{1+b}}
\end{equation}

\begin{figure}
    \centering
    \includegraphics[width=0.4\textwidth]{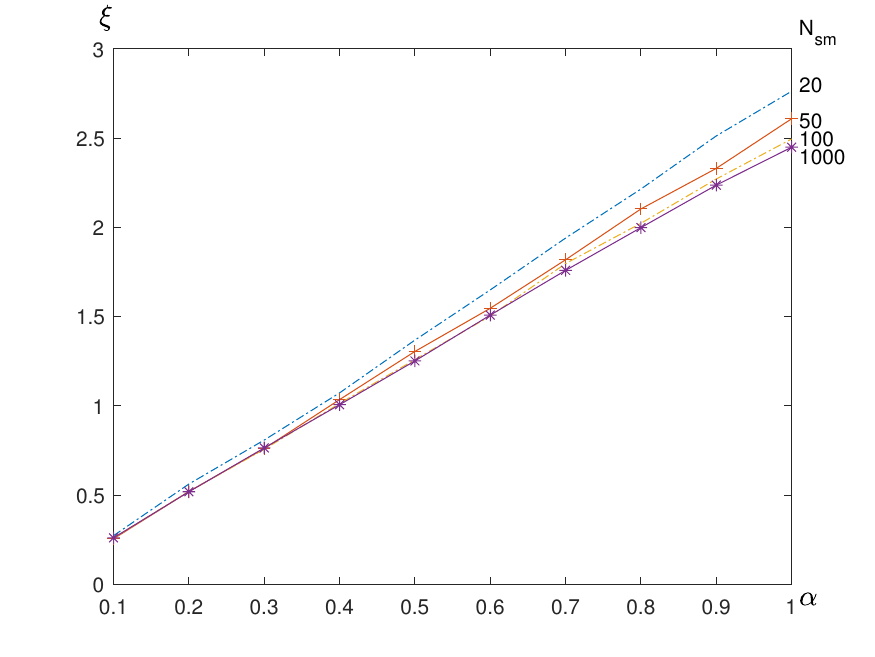}
     \includegraphics[width=0.4\textwidth]{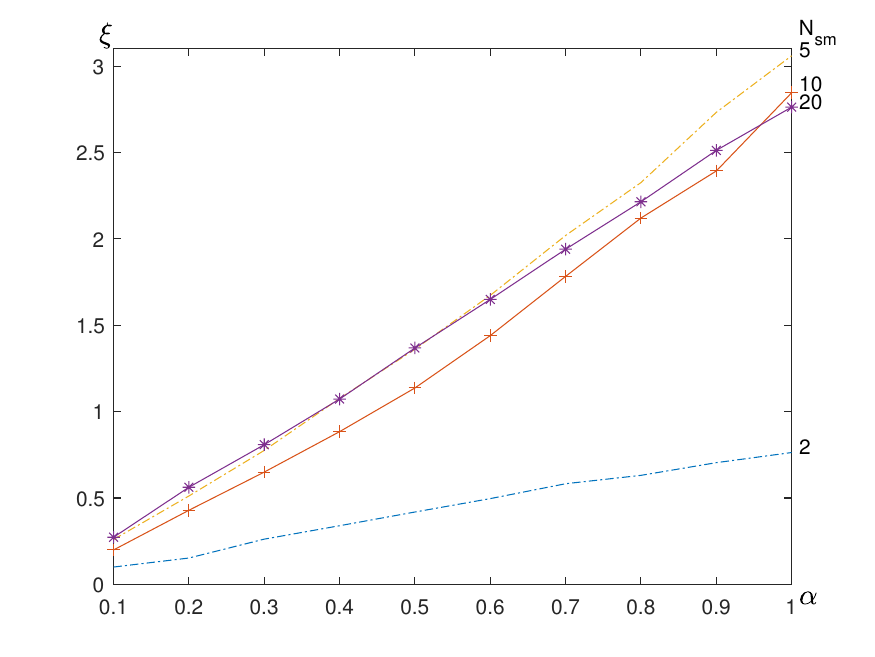}

      \caption{The parameter $\xi$ vs variant $\alpha$ and different $ N_{sm} $, the curves are simple linear functions as parameter $ N_{sm}>10 $.  When $N_{sm}\leq 10$,  $\xi$ can be simply expressed as a formula $\xi=\kappa_\xi \alpha e^{\eta_\xi \alpha}$ . }
    \label{fig: fig12}
\end{figure}

The curves of $\xi$ are simple linear functions as parameter $ N_{sm}>10 $ as shown in Fig.~\ref{fig: fig12}.  When $N_{sm}\leq 10$,  $\xi$ can be simply expressed as a formula
\begin{equation}
 \xi=\kappa_\xi \alpha e^{\eta_\xi \alpha}
\end{equation}

When the flow rate $N_{sm}$ is large, these curves are simple linear($\eta_\xi=0$) to $\alpha$ and the slop $\kappa_\xi\simeq 2.5$.
Therefore, parameters $b$ and $\xi$ can be determined using above simple linear functions of $\alpha$ for  $N_{sm}> 10$. The values of $E_F$ can be obtained using conjectured $b$ and $\xi$ as
 \begin{equation}
 E_F=(1+b)\ln{\frac{\xi-b}{b+1}}
\end{equation}

\begin{figure}
    \centering
    \includegraphics[width=0.4\textwidth]{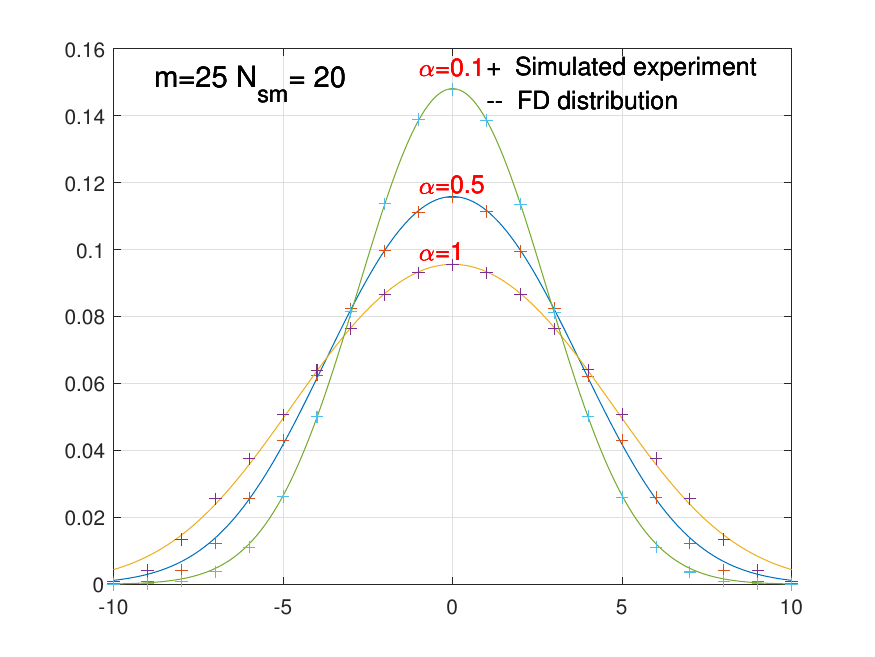}
     \includegraphics[width=0.4\textwidth]{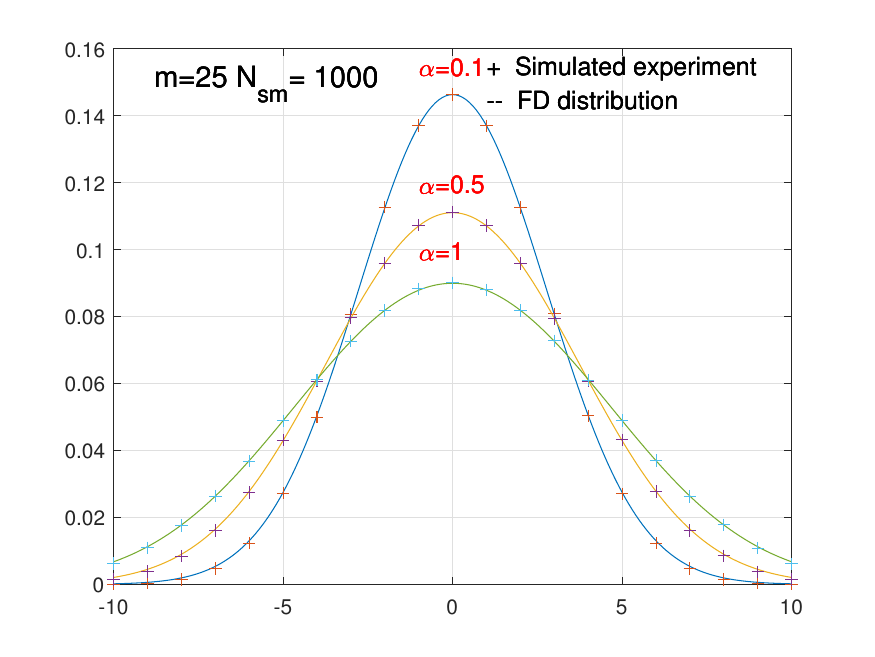}

      \caption{The simple linear relationship conjecture formulas can well fit the experiment when the flow rate  $N_{sm}$ is large. }
    \label{fig: fig13}
\end{figure}
The above simple linear relationship conjecture formulas can well fit the experiment when the flow rate  $N_{sm}>10$ as shown in Fig.~\ref{fig: fig13}.

There are still some fluctuations in the curve when  $ N_{sm}<10 $  , which come from uncertain fitting $E_F$. In order to  determine the slope $\kappa_\xi$ and
exponential factor $\eta_\xi$ when  $ N_{sm} $ is small , we need to eliminate these fluctuations as much as possible. So, we fixed the $b$ parameters using the above conjecture formula, fitted the experiment data  again to obtain the smooth curves of $E_F$ and $\xi$.

\begin{figure}
    \centering
    \includegraphics[width=0.4\textwidth]{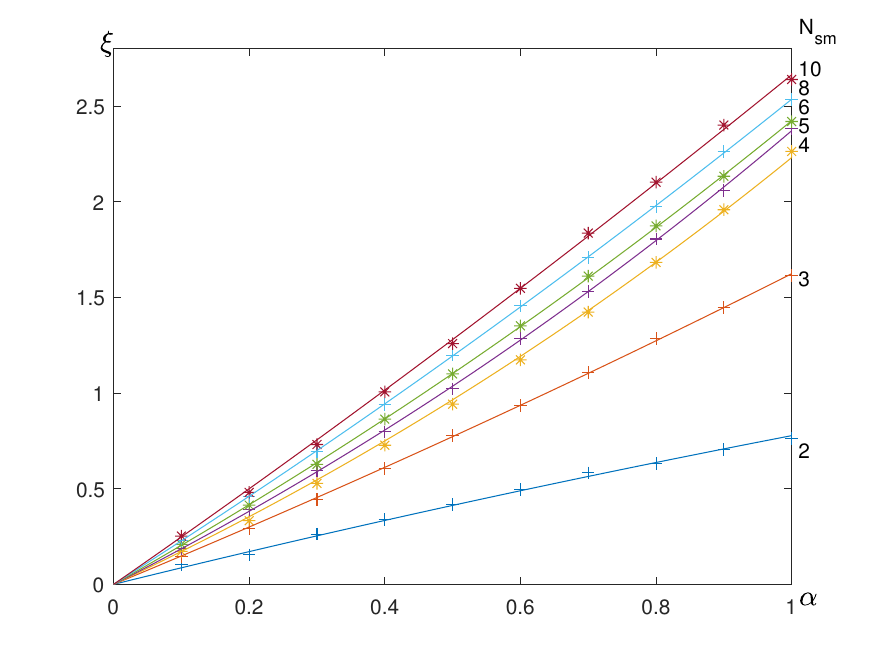}
     \includegraphics[width=0.4\textwidth]{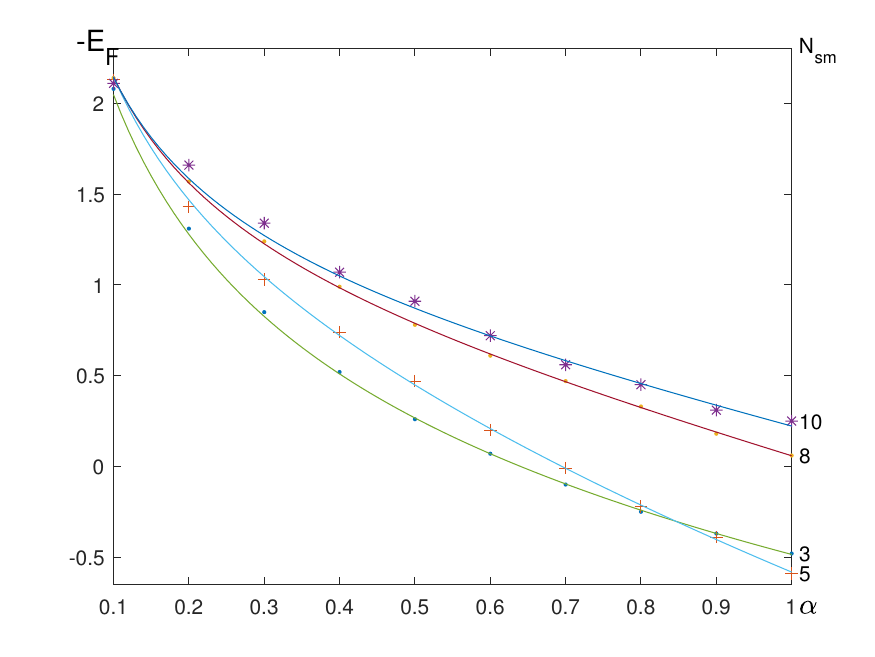}

      \caption{The parameter $\xi$ vs variant $\alpha$ and different small $ N_{sm} $, the $\xi$ curves can be simply expressed as a formula $\xi=\kappa_\xi \alpha e^{\eta_\xi \alpha}$ . The best-fitting $E_F$ also become smooth and conform to the conjecture formula. }
    \label{fig: fig14}
\end{figure}

It can be seen in Fig.~\ref{fig: fig14} that the $\xi$ curves can be simply expressed as a formula $\xi=\kappa_\xi \alpha e^{\eta_\xi \alpha}$ . The slop $\kappa_\xi$ increases with the flow rate $N_{sm}$ from 2 to 10 as the conjecture formula
\begin{equation}
 \kappa_\xi= (2.0+5.66\beta e^{0.12\beta})\kappa_b;\ \ \beta=\frac{1}{N_{sm}-1}
\end{equation}

The exponential factor $\eta_\xi$ is small and changes according to the following polynomial of  $1/N_{sm}$
\begin{equation}
 \eta_\xi= 7.6\beta^2-14.5\beta^3
\end{equation}

\begin{figure}
    \centering
    \includegraphics[width=0.4\textwidth]{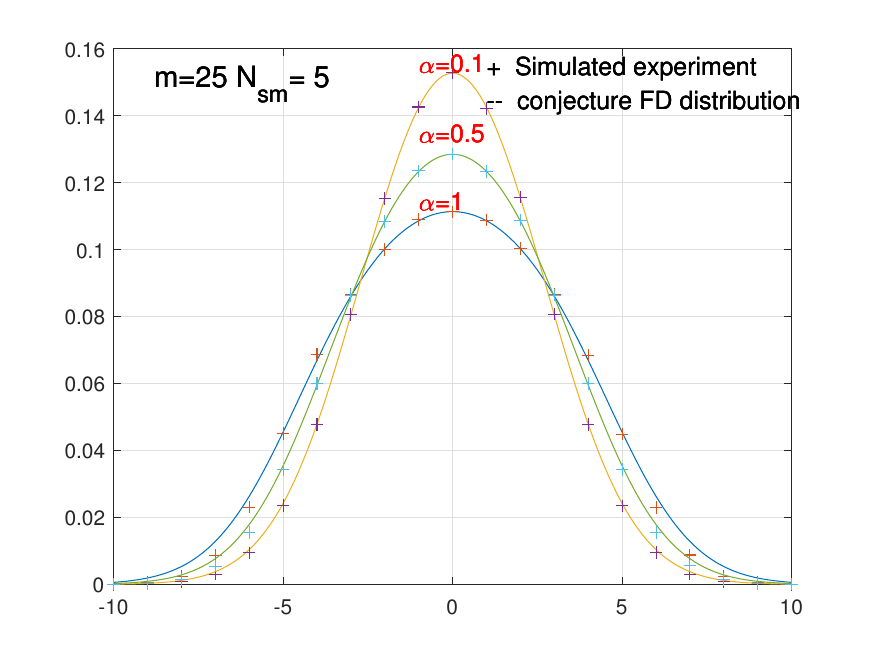}
     \includegraphics[width=0.4\textwidth]{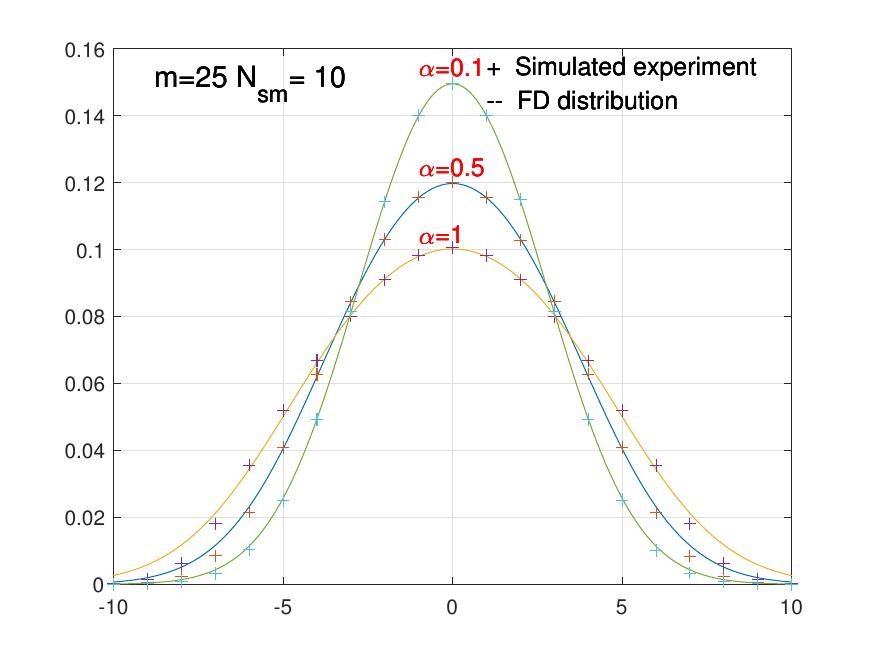}

      \caption{The non-linear conjecture formulas can well fit the experiment when the flow rate  $N_{sm}$ is small. }
    \label{fig: fig15}
\end{figure}

In the Fig.~\ref{fig: fig15}, it can be seen that the above conjecture formula for small  $N_{sm}$  can well fit the experiment data.

Now, we can use a similar conjecture formula to express the distribution of $m=17$.
The parameter $b$ also expressed as:
\begin{equation}
 b=\kappa_b \alpha e^{-\eta_b \alpha};
\end{equation}

The exponential factor $\eta_b$ rapidly decreases to zero with increasing $N_{sm}$
\begin{equation}
 \eta_b= 120\beta^{4};\ \ \beta=\frac{1}{N_{sm}-1}
\end{equation}

When  $N_{sm}>m/2$, the parameter $b$ is still linear to $\alpha$ and the slop $\kappa_b$ slowly increases with the flow rate $N_{sm}$ as the conjecture formula
\begin{equation}
 \kappa_b= 1.68(1-\gamma^2);\ \ \gamma=\frac{m-1}{N_{sm}+m-2}.
\end{equation}

There are only one coefficient changed from $1.99$ for $m=25$ to  $1.68$ for $m=17$ .

When  $N_{sm}>m/2$, the parameter $\xi$ is still linear to $\alpha$ and the slop $\kappa_\xi$ is a constant 2.6 which is a little fine-tuning.

\begin{figure}
    \centering
    \includegraphics[width=0.4\textwidth]{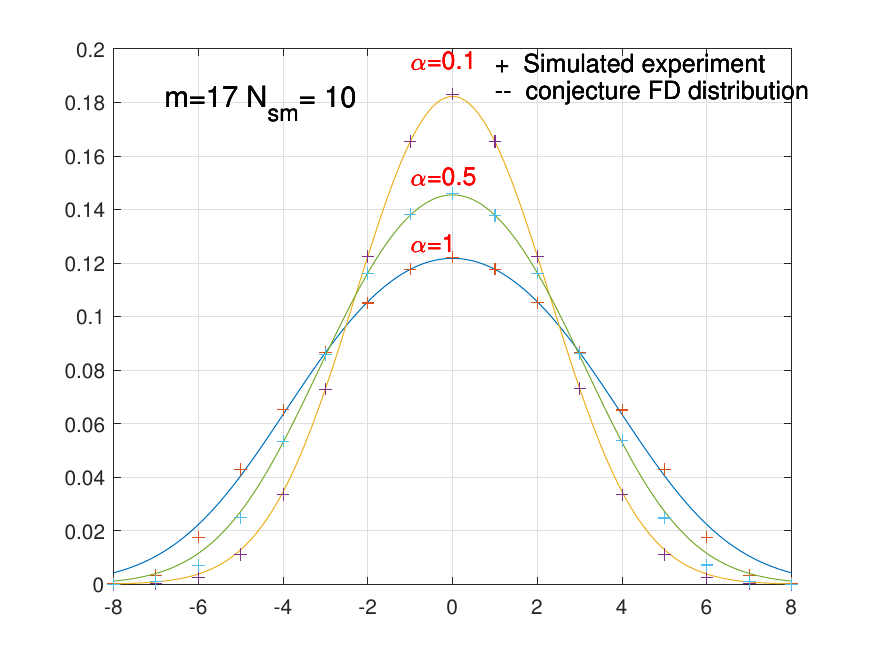}
     \includegraphics[width=0.4\textwidth]{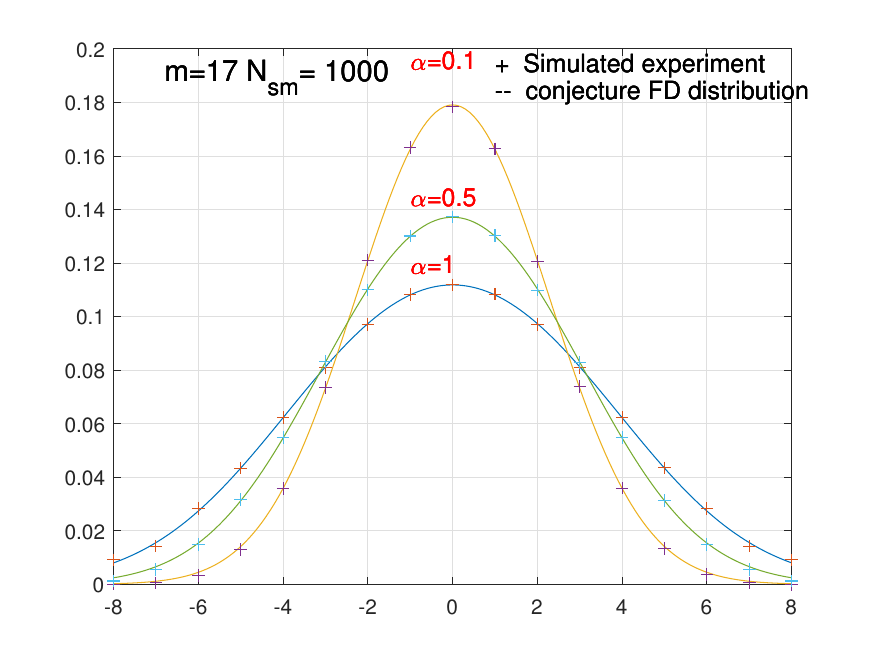}

      \caption{The simple linear relationship conjecture formulas $\xi=2.6\alpha$ and $b= 1.68(1-\gamma^2)\alpha$ can well fit the experiment when the flow rate  $N_{sm}>m/2$ . }
    \label{fig: fig16}
\end{figure}

We fixed the $b$ parameters using the above conjecture formula, fitted the experiment data for $m=17$ again to obtain the smooth curves of $E_F$ and $\xi$.

\begin{figure}
    \centering
    \includegraphics[width=0.4\textwidth]{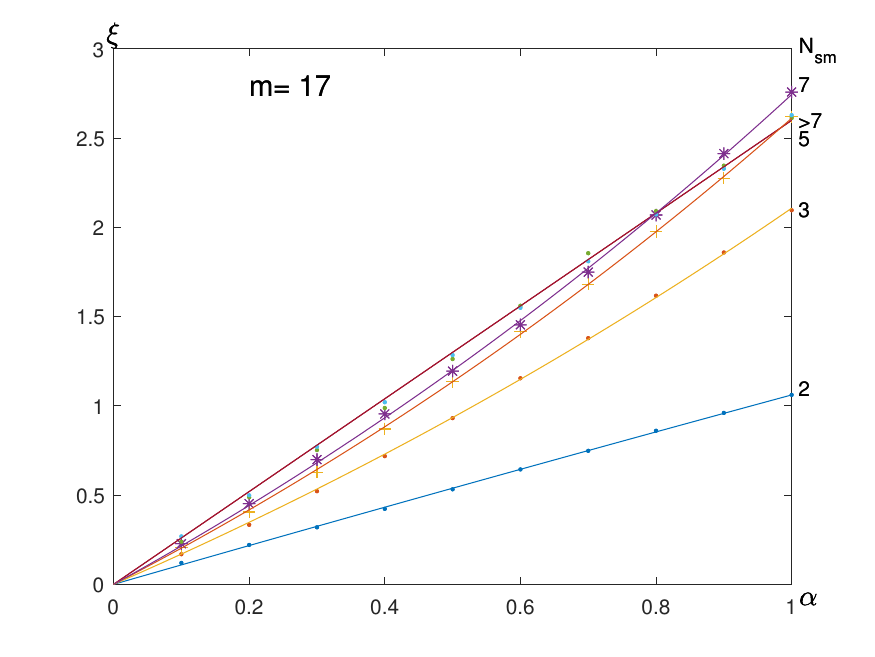}
     \includegraphics[width=0.4\textwidth]{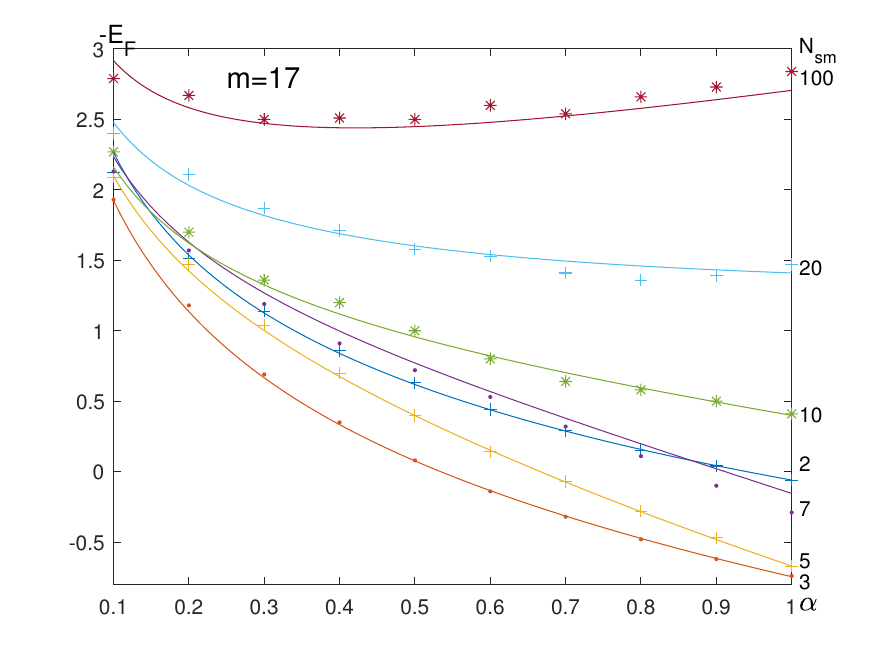}

      \caption{The parameter $\xi$ vs variant $\alpha$ and different small $ N_{sm} $, the $\xi$ curves can be simply expressed as a formula $\xi=\kappa_\xi \alpha e^{\eta_\xi \alpha}$ . The best-fitting $E_F$ also become smooth and conform to the conjecture formula. }
    \label{fig: fig17}
\end{figure}

When  $N_{sm}<m/2$, the $\kappa_b$ become tiny value and the curvature of $\xi$  must be considered as shown in Fig.~\ref{fig: fig17}.  $\xi$ still can be  expressed as
$ \xi=\kappa_\xi \alpha e^{\eta_\xi \alpha}$. Here, $\kappa_\xi$ becomes
\begin{equation}
 \kappa_\xi= (2.17+2.36\beta e^{1.53\beta})\kappa_b;\ \ \beta=\frac{1}{N_{sm}-1}
\end{equation}

The exponential factor $\eta_\xi$  changes according to the following polynomial of  $1/N_{sm}$
\begin{equation}
 \eta_\xi= 12\beta^2-22\beta^3 .
\end{equation}

\begin{figure}
    \centering
    \includegraphics[width=0.4\textwidth]{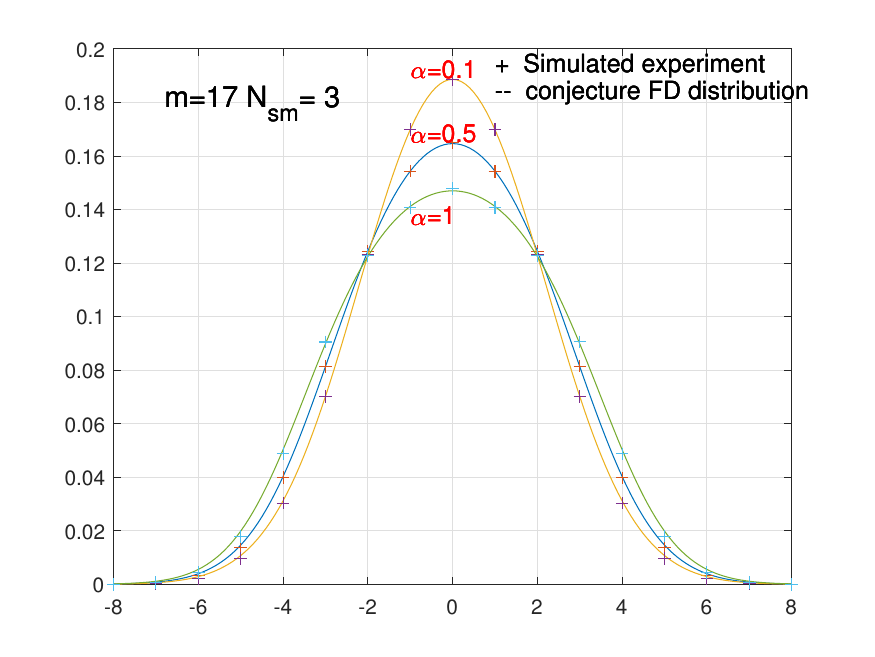}
     \includegraphics[width=0.4\textwidth]{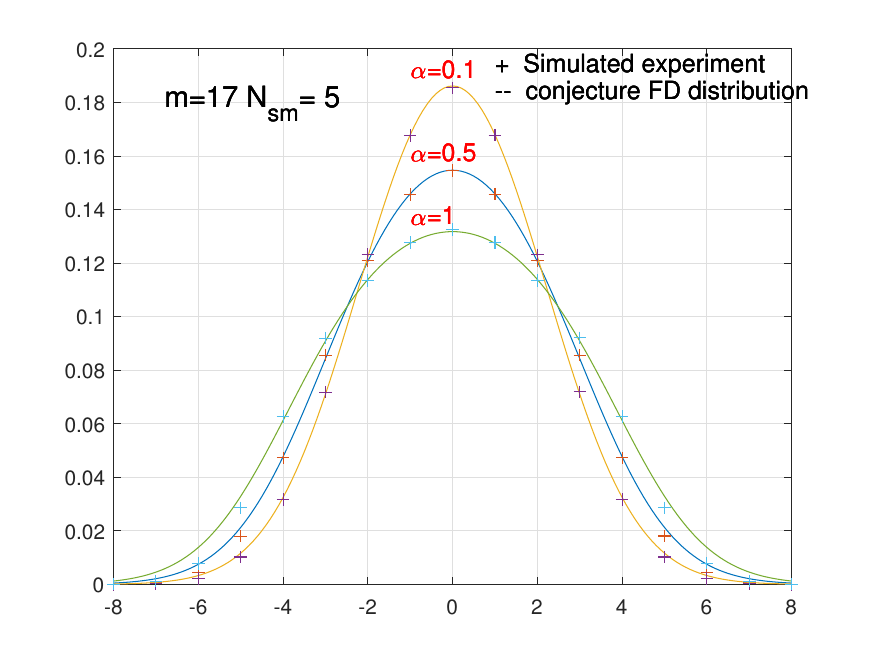}

      \caption{The non-linear conjecture formulas for $m=17$ can well fit the experiment when the flow rate  $N_{sm}$ is small. }
    \label{fig: fig18}
\end{figure}

In Fig.~\ref{fig: fig18}, it can be seen that the above conjecture formulas for small  $N_{sm}$  can well fit the experiment data $m=17$.

The Fermi Dirac distribution will degenerate into a Gaussian distribution function when $\frac{-E_F}{1+b}>2.5$.
\begin{figure}
    \centering
    \includegraphics[width=0.40\textwidth]{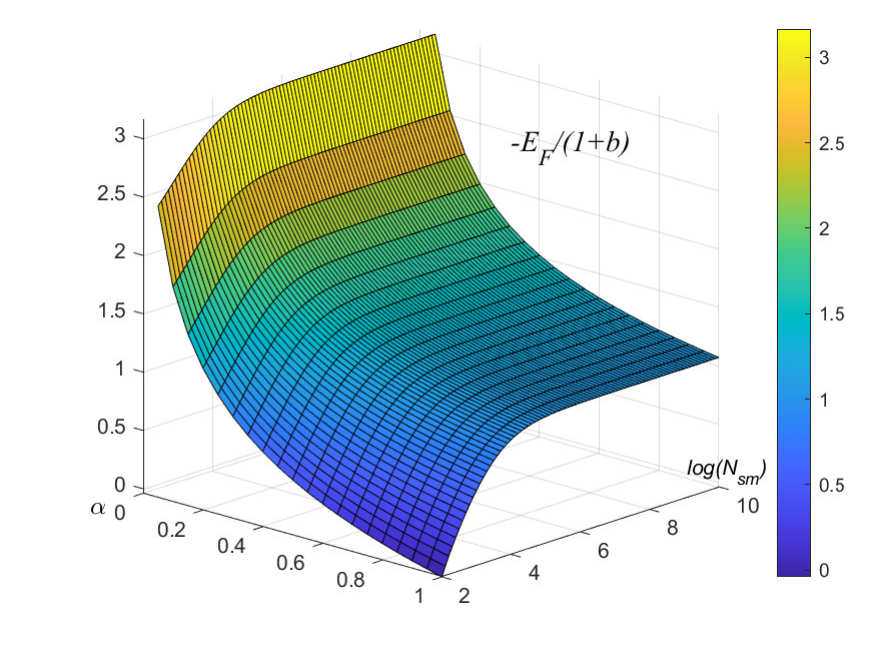}

      \caption{The figure demonstrates the dependence of  $-E_F/(1+b)$ on experimental parameters $\alpha $ and $N_{sm}$ for $m=17$. The Fermi Dirac distribution will degenerate into a Gaussian distribution function when $\alpha$ is small. The upper limit on  $\alpha$ for degradation will increases with $N_{sm}$. }
    \label{fig: fig19}
\end{figure}

The Fig.~\ref{fig: fig19} demonstrates the dependence of $-E_F/(1+b)$ on experimental parameters $\alpha $ and $N_{sm}$ for $m=17$. The Fermi Dirac distribution will degenerate into a Gaussian distribution function when $\alpha$ is small. The upper limit on  $\alpha$ for degradation will increases with $N_{sm}$. When  $N_{sm}$ reaches 1000, The distribution will degenerate as $\alpha<0.3$.

\begin{figure}
    \centering
    \includegraphics[width=0.40\textwidth]{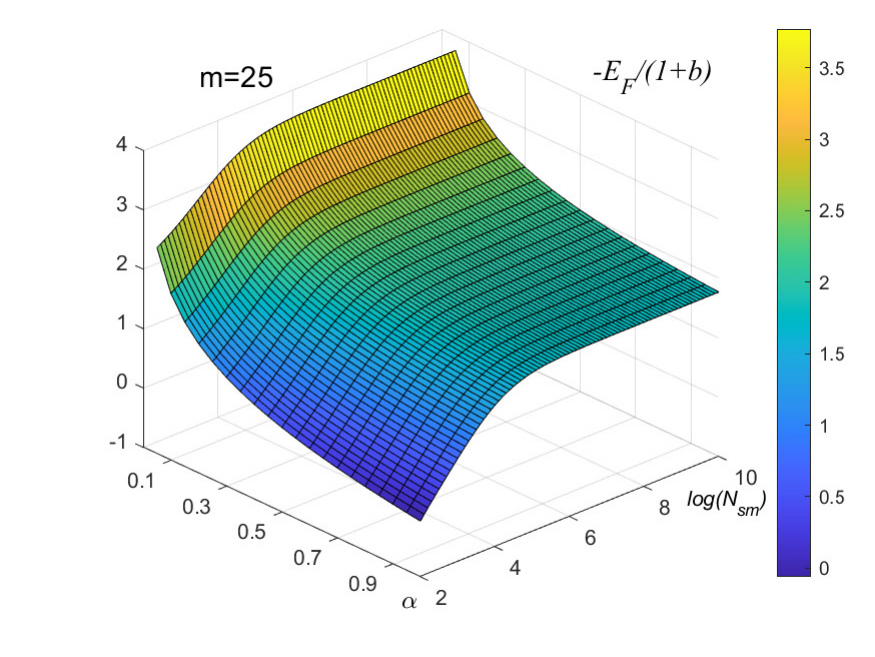}

      \caption{When the value of $m=25$, $-E_F/(1+b)$ will increase. The upper limit on  $\alpha$ for degradation even becomes 0.5 when $N_{sm}$ is large. }
    \label{fig: fig20}
\end{figure}

 The increase of the sluts number $m$ leads to the growth of $-E_F/(1+b)$, making degradation more likely to occur.   It can been seen from Fig.~\ref{fig: fig20} that the distribution even degenerates when $\alpha<0.5$ if $m$ increased to 25. This phenomenon indicates that $m$ represents the number of particle states, and when the number of states is large, the states is almost continuous spectrum. The quantum distribution will degenerate to the Classical limit, and the Fermi Dirac distribution will become the Classical Boltzmann distribution which is equivalent to Gaussian distribution.
  In order to verify the Classical limit of large $m$, we directly extend m to 101. The conjecture formula is still valid, just adjust the $\kappa_b$ coefficient.
  \begin{equation}
 \kappa_b= 2.4(1-\gamma^2);\ \ \gamma=\frac{m-1}{N_{sm}+m-2}.
\end{equation}
while $m=101$.

Meanwhile, when  $N_{sm}>20$, the parameter $\xi$ is still linear to $\alpha$ and the slop $\kappa_\xi$ is a constant 2.5 which is same with $m=25$;

  \begin{figure}
    \centering
    \includegraphics[width=0.40\textwidth]{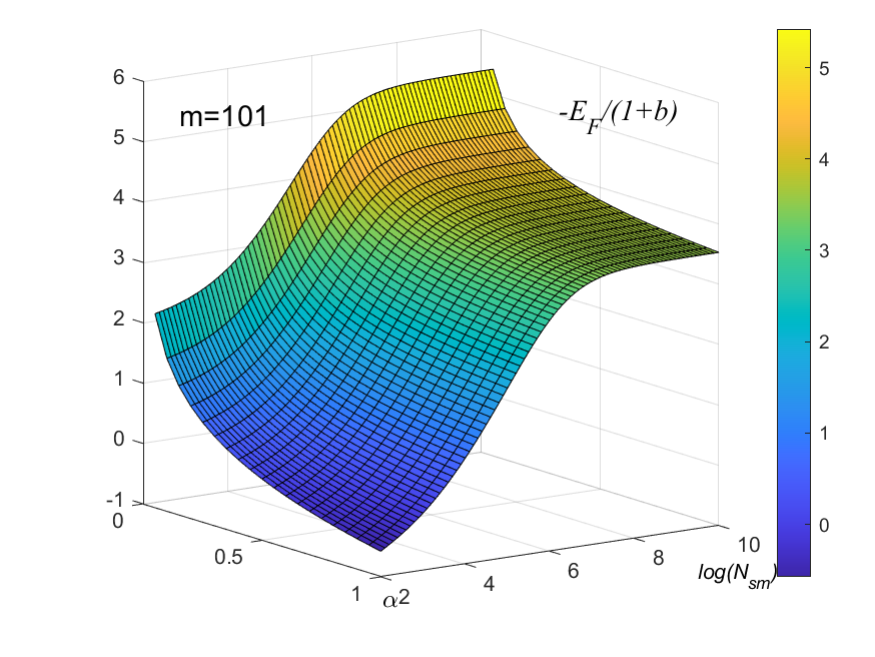}

      \caption{When $m=101$, the value of $-E_F/(1+b)$ is large in most range. Regardless of the magnitude of the interaction, the distribution function can degenerate into a Gaussian distribution when $N_{sm}>100$. }
    \label{fig: fig21}
\end{figure}

The large values of $-E_F/(1+b)$ in Fig.~\ref{fig: fig21} indicate that classical correspondence does occur in the case of  continuous spectrum.  The distribution function returns to a broadened Gaussian distribution. And the broadening ratio of $sigma^2$ is a simple linear relationship with interaction factor $\alpha$.

\section{Real Galton Board Experiment}
By now, all the study is based on simulation experiment data. In order to verify whether the above analysis results match the real Galton Board experiment, a real instrument with m=17 is used to obtain real data. It is difficult to measure the interaction  intensity of particles in real  Galton Board. The flow rate of particles falling simultaneously is also difficult to accurately control. So the values of $\alpha$ and $N_{sm}$  can only be determined based on experimental data.

\begin{figure}
    \centering
    \includegraphics[width=0.40\textwidth]{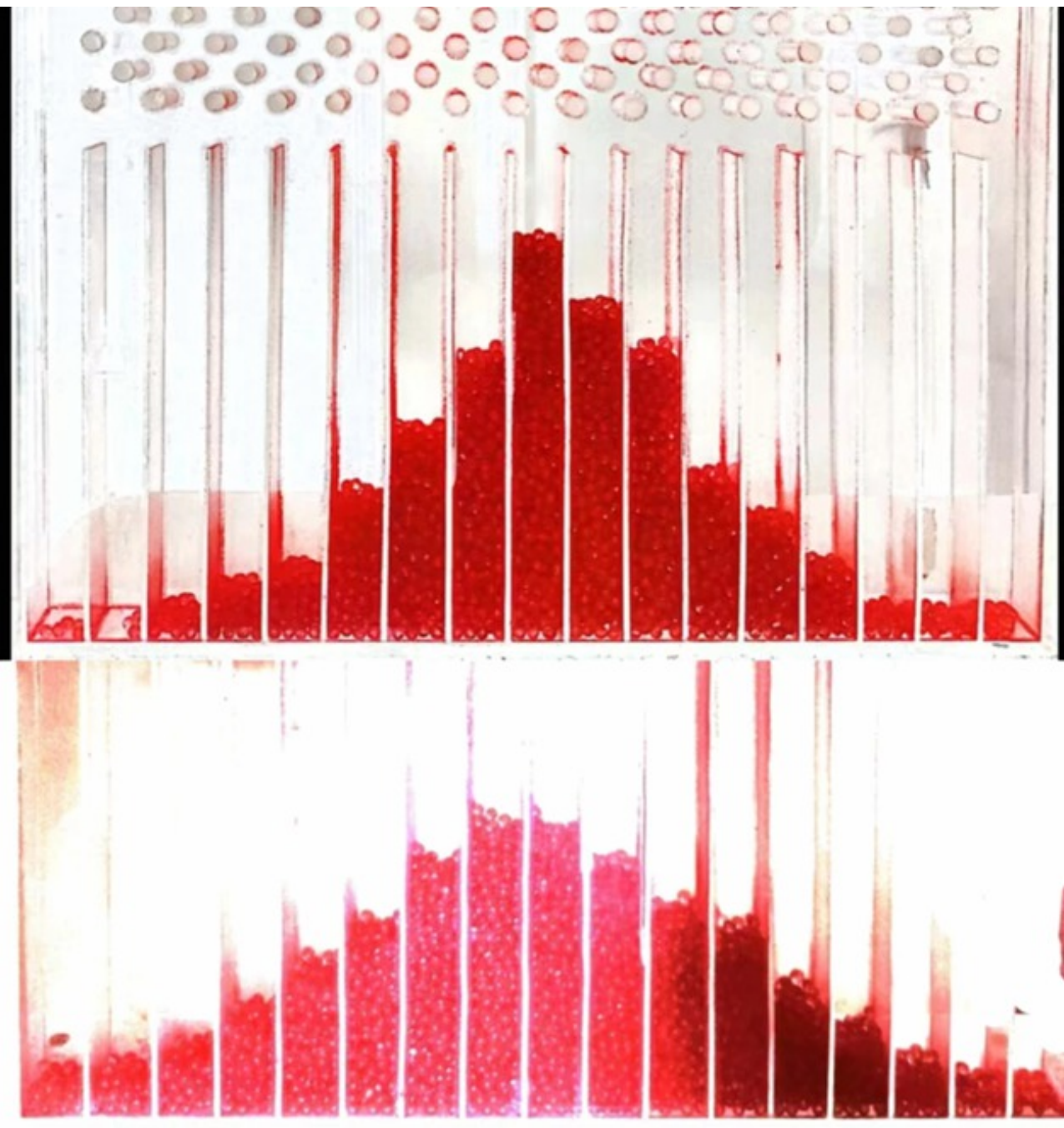}

      \caption{The above figure shows the distribution of small flow real experiments and  the following figure shows the high flow. }
    \label{fig: fig22}
\end{figure}

The real Galton Board is much more complex than ideal model. Particles falling alone do not follow a binomial distribution, because the step size of particle movement is not fixed $\pm 0.5$. The Gaussian distribution is still regarded as the probability distribution function when  particles fall alone, but the standard deviation $\sigma \neq 2.02$ and needs to be measured. We first use small particles flow to obtain the standard deviation. The true standard deviation is 2.3 measured from the particle distribution data in Fig.~\ref{fig: fig22}.

\begin{figure}
    \centering
    \includegraphics[width=0.40\textwidth]{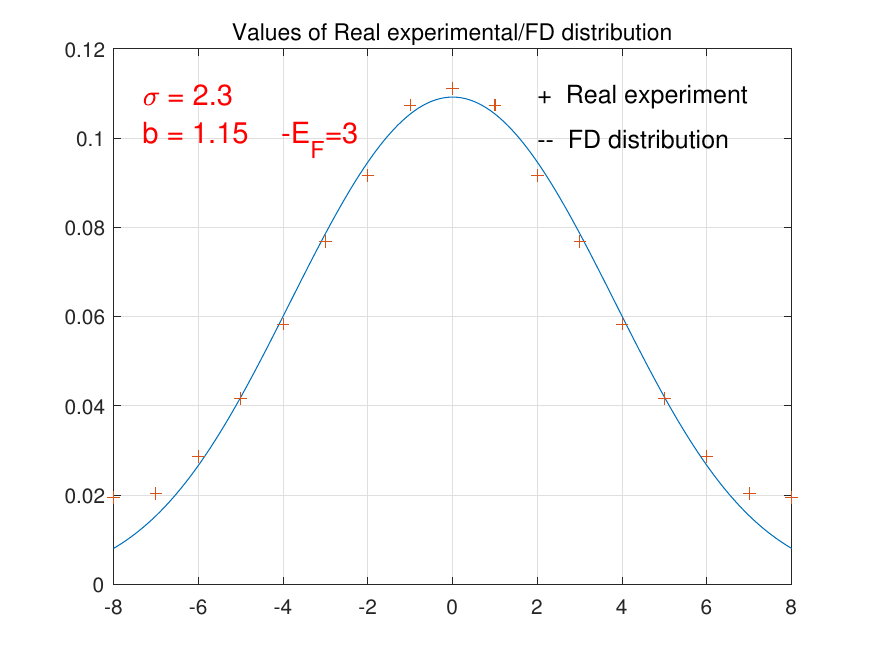}

      \caption{ The Fermi Dirac distribution function with $b=1.15$ and $-E_F=3$ is consistent with the real experimental distribution data. }
    \label{fig: fig23}
\end{figure}

Real experimental devices can only control the flow rate within a rough range, so  $N_{sm}$   is not fixed. The particles interaction factor $\alpha$ also cannot be determined as a constant independently of  $N_{sm}$  , just like the ideal model. Therefore, we can only roughly determine the interaction factor $\alpha$ corresponding to the flow rate  $N_{sm}$  within a certain range from the experimental results.

   The basic standard deviation measured for this experimental device is $\sigma=2.3$, this standard deviation corresponds to the value $m=22$ in the ideal model.According to the conclusion of the simulation experiment,  the parameter $b$ is linear to $\alpha$, When $N_{sm}>m/2$, and the slop $\kappa_b$ slowly increases with the flow rate $N_{sm}$ as the conjecture formula for $m=22$
\begin{equation}
 \kappa_b= 1.9(1-\gamma^2);\ \ \gamma=\frac{m-1}{N_{sm}+m-2}.
\end{equation}

Based on the real experimental results of Fig.~\ref{fig: fig23}, the flow rate $N_{sm}$ is approximately 20-50. According to the conjecture formula of the simulation experiment, the equivalent interaction factor of this device $\alpha\simeq 0.6-0.7$ when $N_{sm}\simeq 20-50$.

\section{Attraction and Bose-Einstein Distribution}

\begin{figure}
    \centering
    \includegraphics[width=0.4\textwidth]{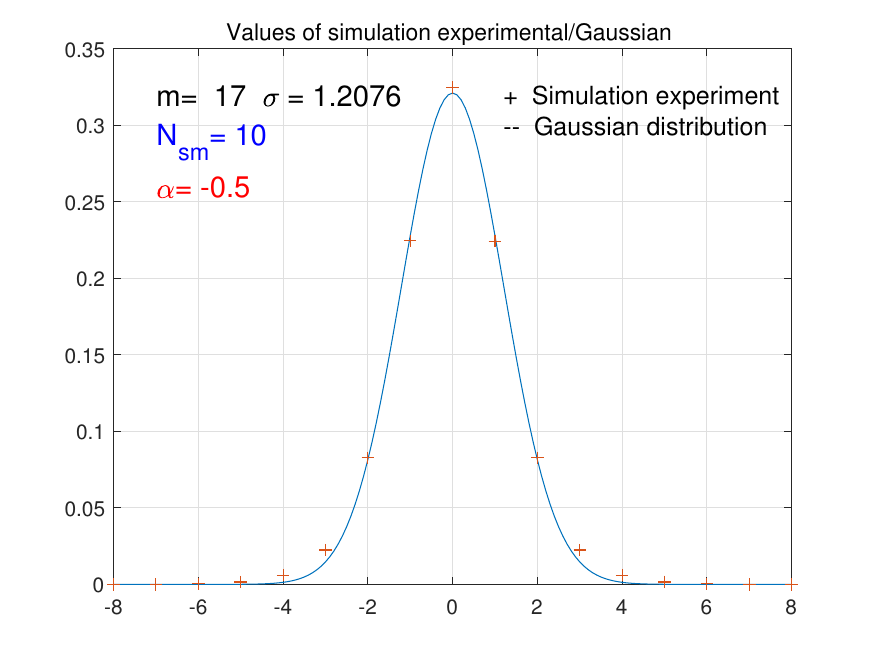}
     \includegraphics[width=0.4\textwidth]{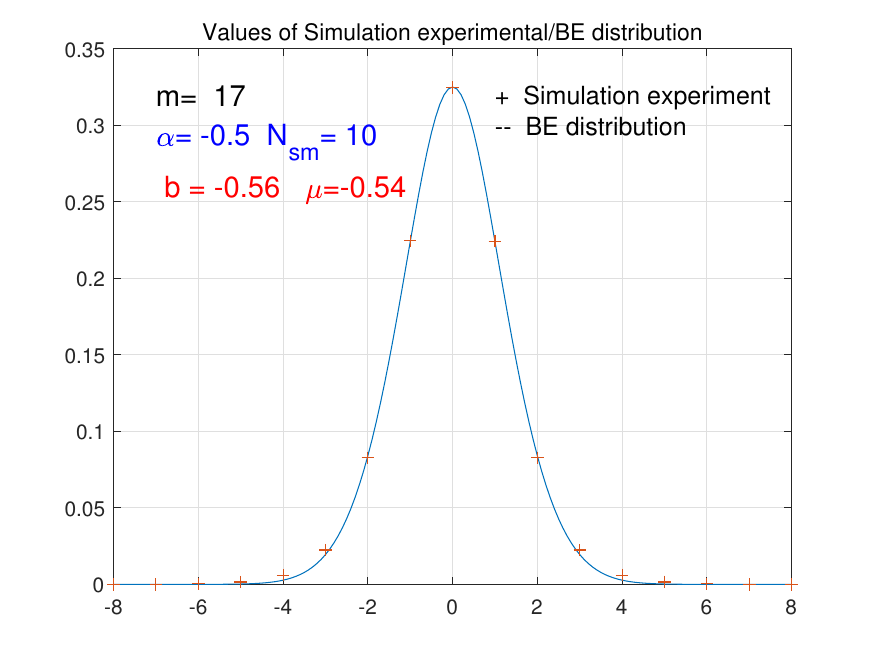}

      \caption{The Bose-Einstein distribution function(Right) fits the experiment much better than Gaussian distribution(Left)}
    \label{fig: fig24}
\end{figure}

The interesting results of simulation experiments of Galton boards with exclusivity between particles can be easily extended to other hypothetical models. It is possible to imagine the situation where particles have attraction interaction. We can conveniently set $\alpha$  to negative values to parameterize the Toy model. Like the exclusive Galton board, when the attraction is in the middle, the probability distribution deviates from the Gaussian distribution which can been seen in Fig.~\ref{fig: fig24}(Left). It's reasonable to assume that particles distribution law  would be Bose-Einstein distribution functions. From Fig.~\ref{fig: fig24}(Right), it can be seen that the Bose-Einstein distribution function fits the experiment much better.

We continue to use $1+b$ as the temperature parameter in the distribution functions, and $\mu$ is the Chemical potential. Attraction will cause $b$ to be negative, while its value close to -1 representing the appearance of Bose-Einstein condensation.

\begin{equation}\label{eq: BE_equation}
f(x)\propto \frac {1}{e^ {\frac {E-\mu}{1+b}}-1};\  E= \frac {x^2}{2\sigma^2}.
\end{equation}

\begin{figure}
    \centering
    \includegraphics[width=0.4\textwidth]{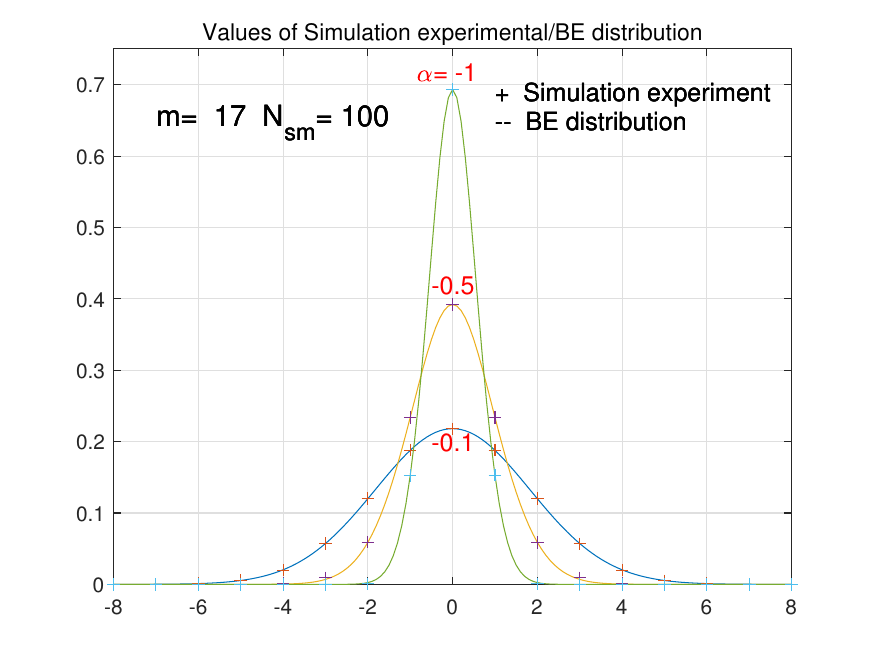}
     \includegraphics[width=0.4\textwidth]{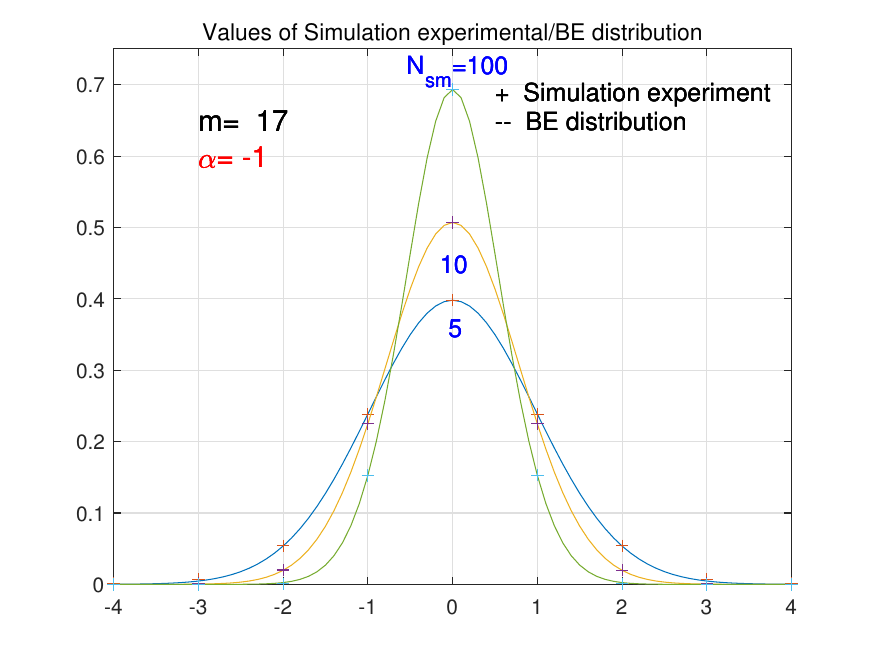}

      \caption{The distribution width $1+b$ decreases with the increase of $-\alpha$ and  $N_{sm}$, indicating that attraction and particle density enhance Bose Einstein condensation }
    \label{fig: fig25}
\end{figure}

The distribution width $1+b$ is very narrow when attraction $-\alpha$ is strong and  particle number  $N_{sm}$ is large as shown in Fig.~\ref{fig: fig25}, meaning almost particles condense into the same state.

\begin{figure}
    \centering
    \includegraphics[width=0.4\textwidth]{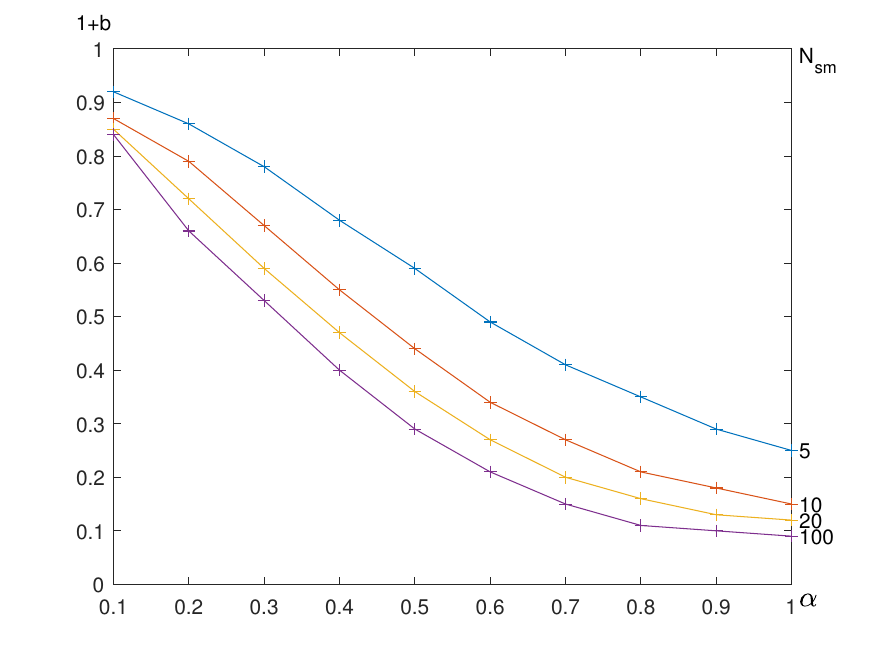}
     \includegraphics[width=0.4\textwidth]{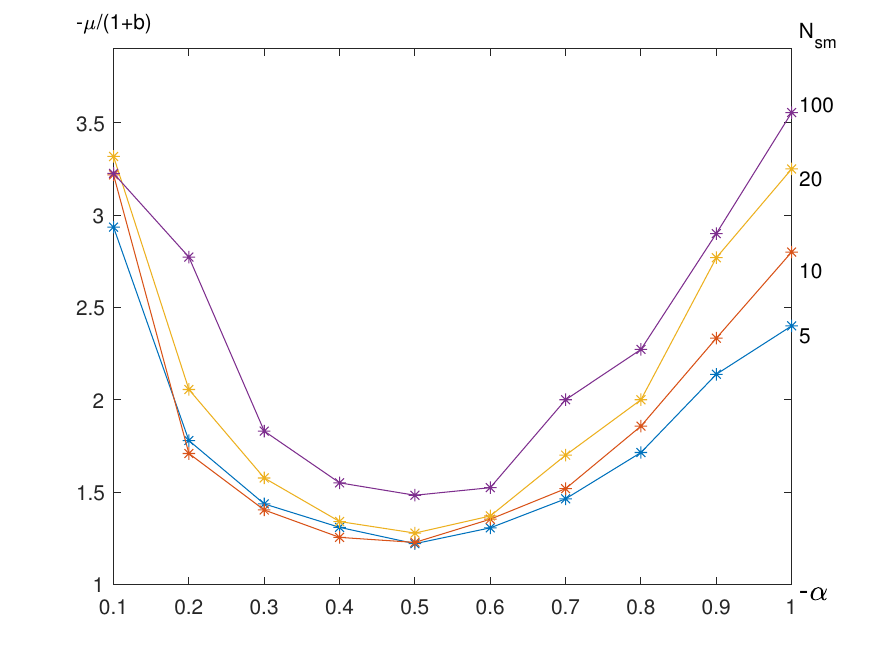}

      \caption{The distribution width $1+b$ decreases with $-\alpha$, and  $-\alpha/(1+b)$ is not large.}
    \label{fig: fig26}
\end{figure}

The distribution square-width $1+b$(also the temperature parameter) decreases with $-\alpha$ rapidly and linearly  until $1+b<0.2$ which meaning the particles are only distributed in three positions. The parameter Chemical potential $-\mu$ over "temperature"($kT=1+b$) represents the difference between quantum distribution and classical distribution functions. From Fig.~\ref{fig: fig26}(Right), it can be seen that Bose-Einstein distributions cannot degenerate into  Gaussian distributions when $-\alpha$ is in the middle value.

\section{Discussion and Conclusions}
\label{sec: discussion}
Through simulation and real experiments, it was found that the distribution of particles in Galton Boards cannot be simply represented by a Gaussian distribution or even a binomial distribution. When the number of slots $m$ is not too large, the particle's position states is discrete. And when the flow rate $N_{sm}$ is not very large, the particles appear as a whole at a certain position. The wholeness and exclusivity of particles result in the distribution law following the Fermi Dirac function, reflecting quantum effects.  When the flow  $N_{sm}$  value is maximum, there are enough particles that can be distributed to different positions according to probability distribution, which is equivalent to particles being able to be divided. Therefor returning to the classical distribution. This simple model can simulate the diffusion process of exclusive Fermion, such as electron gas in metals.

This simple toy model can also simulate Bose-Einstein condensation, just when the parameter $\alpha$ are adjusted to negative for attractions. In the model, the number of layers/slots $m$ represents the number of particle states or the evolution time of diffusion, the number of particles falling simultaneously $N_{sm}$ represents the total particle number in the system. As m and n increase, the particle distribution exhibits a simple statistical pattern. At the same time, when the interaction strength is not too large, the distribution parameters exhibit a linear dependence on the strength. In the extreme case of large numbers, the distribution function and its parameters can be obtained by solving the functional extremum  to obtain the distribution function with maximum entropy\cite{BGG}.

\end{document}